\documentclass[12pt]{article}

\usepackage[english]{babel}

\usepackage[letterpaper,top=2cm,bottom=2cm,left=1.5cm,right=1.5cm,marginparwidth=1.75cm]{geometry}

\usepackage{amsmath,amssymb,amsmath,mathtools,eepic}
\usepackage{xcolor}
\usepackage{graphicx,psfrag,pict2e}
\usepackage[colorlinks=true, linkcolor=blue, citecolor=red]{hyperref}

\usepackage{tikz-cd}
\usepackage{subcaption}

\usepackage{csquotes}

\usepackage[backend=biber,style=alphabetic,sorting=ynt]{biblatex}
\bibliography{references.bib}

\renewcommand{\bar}{\overline}

\DeclareMathOperator{\tr}{tr}
\DeclareMathOperator{\Lie}{Lie}
\DeclareMathOperator{\Ad}{Ad}

\DeclareMathOperator{\End}{End}
\DeclareMathOperator{\Aut}{Aut}
\DeclareMathOperator{\Out}{Out}

\newcommand{\R}{\mathcal{R}}
\newcommand{\id}{\mathrm{id}}
\newcommand{\GG}{\mathrm{G}}
\newcommand{\HH}{\mathrm{H}}
\newcommand{\OO}{\mathrm{O}}
\newcommand{\SO}{\mathrm{SO}}
\newcommand{\so}{\mathfrak{so}}
\newcommand{\RR}{\mathcal{R}}
\renewcommand{\SS}{\mathbb{S}}
\renewcommand{\gg}{\mathfrak{g}}
\newcommand{\hh}{\mathfrak{h}}
\newcommand{\mm}{\mathfrak{m}}

\numberwithin{equation}{section}

\title{Scattering in the dual regime of\\
Yang-Baxter deformed $\mathrm{O}(2N)$ sigma models}
\author{Alexey Bychkov$^{1,2,3}$ and  Boris Nekrasov$^{1}$
\vspace{0.5cm}
\\
{\normalsize\it\centerline{1. Skolkovo Institute of Science and Technology, 121205 Moscow, Russia}}
\\
{\normalsize\it\centerline{2. HSE University, 6 Usacheva str., Moscow 119048, Russia}}\\
{\normalsize\it\centerline{3. Moscow Institute of Physics and Technology, 141700 Dolgoprudny, Russia}}
}

\date{}

\begin{document}
\maketitle
\begin{abstract}
    We continue to explore the previously suggested dual regime of Yang-Baxter (YB) deformed $\OO(2N)$ sigma models, which is a new one-parametric deformation of the $\OO(2N)$ model. It can be obtained from the conventional YB deformed $\OO(2N+2)$ sigma model by freezing two isometries. The scattering matrix in the non-deformed $\OO(n)$ model is known to coincide with the rational $\mathfrak{so}_n$-invariant $R-$matrix. For $n = 2N$ this rational solution allows for two trigonometric deformations: the usual $D_N^{(1)}$ and the twisted $D_{N}^{(2)}$. The usual one is known to coincide with the scattering matrix of the conventional YB-$\OO(2N)$ model. A natural question to ask is what sigma model corresponds to the twisted solution. In this work we claim that it is precisely our dual regime of YB-$\OO(2N)$. We find the corresponding unitarizing function $F_N(\theta)$ explicitly, and, as a bonus, extract some physical properties of this system using the thermodynamic Bethe Ansatz technique.
\end{abstract}

\tableofcontents
\section{Introduction}
Two-dimensional nonlinear sigma models \cite{Gell-Mann:1960mvl} provide toy models for realistic four-dimensional interacting quantum field theories (QFTs) sharing a number of qualitative phenomena with them, such as asymptotic freedom, instantons and renormalizability \cite{Friedan:1980jm}. Some of these models are special, being in principle exactly soluble on both classical and quantum levels. A number of such integrable models is known: principal chiral model (PCM), in which fundamental fields take values in a Lie group $\GG$; Wess-Zumino-Novikov-Witten (WZNW or WZW) models, which are PCMs plus non-trivial topological $B-$field;  symmetric space sigma models (SSSM) with target space being symmetric coset $\GG/\HH$; gauged WZW models, to name a few. Classical integrability in these theories is usually established by constructing a Lax connection: if a system allows for a current which is both flat and conserved on shell, then a Zakharov-Mikhailov mechanism shows existence of an infinite tower of conserved charges. These ideas were applied for PCM and SSSM in \cite{Zakharov:1973pp, Eichenherr:1979ci}. Generalizations for WZW and gauged WZW models are possible as well. We direct an interested reader to an excellent pedagogical review \cite{Hoare:2021dix}.

PCM, PCM+WZW, SSSM and gauged WZW models, of course, do not comprise an exhausting list of integrable sigma models. For example, a very important class of integrable models are one-parametric deformations of them that preserve Lax integrability and 1-loop RG-stability. Importantly, in all known instances classical integrability appears in a such theories along with RG-stability (Ricci flow reduces to ODE) and quantum integrability (factorized scattering). It is thus generally believed that the classical property of Lax integrability, geometric property of RG-stability and quantum property of factorized scattering are indeed connected, presumably via some large symmetry, a quantum group.

Let us concentrate for a moment on factorized scattering. In a seminal paper \cite{Zamolodchikov:1978xm} the general defining properties of $S-$matrices in integrable theories were established: factorization, unitarity and crossing symmetry. More explicitly, it was shown that integrability (infinite number of conserved charges) manifests itself in the factorization of general $n\rightarrow n$ scattering amplitude down to a product of elementary $2\rightarrow2$ ones. In this way, the natural object of interest in integrable QFT is a $2\rightarrow 2$ scattering matrix $S_{12}(\theta)$. It depends on the difference in rapidities of colliding particles and carries the information about their internal degrees of freedom, so that $S-$matrix becomes a $\theta$-dependent operator acting on the tensor product of two single-particle state spaces: $S_{12}(\theta)\in\End(\mathcal H_1\otimes\mathcal H_2)$. Factorization property is then expressed by a celebrated quantum Yang-Baxter equation (YBE):
\begin{equation}\label{quantum-YBE}
    S_{12}(\theta_1 - \theta_2)S_{13}(\theta_1 - \theta_3)S_{23}(\theta_2 - \theta_3) = S_{23}(\theta_2 - \theta_3) S_{13}(\theta_1 - \theta_3) S_{12}(\theta_1 - \theta_2).
\end{equation}
Its solutions are called quantum $R-$matrices and are studied very thoroughly in theoretical physics and in pure mathematics. Some of the classification results will be important for us in this work. 

Unitarity and crossing symmetry conditions are, in contrast, the physical ones. They read respectively
\begin{align}\label{unitarity&crossing}
    &S_{12}(\theta)S_{21}(-\theta) = \mathrm{id},\\
    &S_{12}(\theta) = C_1 S_{2\tilde{1}}(i\pi - \theta) C_1^{-1},
\end{align}
where $\tilde 1$ means transposition in $\mathcal{H}_{1}$ and $C_1\in\End(\mathcal H_1)$ is a charge conjugation matrix. Equations \eqref{quantum-YBE}, \eqref{unitarity&crossing} are called bootstrap equations and are extremely constraining. Along with analyticity in $\theta$ and possibly some global symmetry they allow for $S_{12}(\theta)$ to be found explicitly. 

This program was first successfully implemented in \cite{Zamolodchikov:1978xm} for the situation with global $\OO(N)$ symmetry, where exact solutions to the bootstrap equations were found for every $N\geq 3$. The dependence on the rapidity $\theta$ turned out to be rational, hence the name: rational $R-$matrix. Let us briefly stress one important feature of this construction. Let $R_{12}(\theta)$ be any analytic solution to YBE \eqref{quantum-YBE}. Then for any analytic function $F(\theta)$ the tensor $F(\theta)R_{12}(\theta)$ is a solution as well. This function $F(\theta)$ is usually referred to as unitarizing factor, since unitarity and crossing equations for $S_{12}(\theta) = F(\theta) R_{12}(\theta)$ essentially boil down to two functional equations for $F(\theta)$. By solving these equations it is possible to make any quantum $R-$matrix, which a priori satisfies only YBE, into a scattering matrix of some integrable theory, which solves all three of the bootstrap equations.\footnote{In fact, for a fixed $R-$matrix $R_{12}(\theta)$ there are typically many solutions for $F(\theta)$. For instance, in \cite{Zamolodchikov:1978xm} a single $\OO(N)$-invariant R-matrix equipped with two different unitarizing factors $F(\theta)$ was shown to describe scattering in two different theories: in a round sphere sigma model ($\frac{\SO(N)}{\SO(N-1)}$ SSSM) and in $\OO(N)$ Gross-Neveu model respectively.}

There is, however, much more to this story: the trigonometric deformations. These are families of quantum $R-$matrices depending on rapidities as $f(e^{\lambda\theta})$, where $\lambda$ is a deformation parameter and $f$ is a rational function (hence the name: trigonometric), and such that the limit $\lambda\rightarrow 0$ produces the rational solution (hence the name: deformation). It is well known that for any complex finite-dimensional simple Lie algebra $\gg$ trigonometric deformations exist  \cite{Belavin1998TriangleEA,Bazhanov:1984gu,Jimbo:1985ua} and comprise discrete series depending on some combinatorial data (which is now called Belavin-Drindeld triples). Moreover, upon fixing this combinatorial data, the trigonometric deformation of the $r$-matrix is unique for $\gg=B_n$ and $C_n$, whereas for series $A_n$ and $D_n$ there are two distinct ones: the regular and the twisted, which are usually called type $\gg_n^{(1)}$ and type $\gg_n^{(2)}$ respectively, where $\gg_n = A_n, D_n$. The existence of a twisted solution of type $ D_n^{(2)} \simeq {\so}_{2n}^{(2)}$ is of utmost importance for us. 

A simplest nontrivial round sphere sigma model is $\OO(3)$ theory\footnote{Following the tradition, by $\OO(N)$ model we always mean the $\frac{\SO(N)}{\SO(N-1)}$ symmetric space sigma model (SSSM) and \textit{not} the PCM with target space being $\OO(N)$ group.}. One could ask how does the trigonometric deformation change the action of the corresponding sigma model. In a seminal paper \cite{Fateev:1992tk} an explicit answer was given: the round 2-sphere deforms to a so-called sausage:
\begin{equation}
    \begin{aligned}
        ds^2 \sim d\theta^2 + \cos^2\theta\,d\phi^2 &\quad\xrightarrow{\text{trig.def.}}\quad ds^2 \sim\frac{d\theta^2 + \cos^2\theta\,d\phi^2}{1-\kappa^2\sin^2\theta},\\
        R_\text{rat.} &\quad\xrightarrow{\text{trig.def.}}\quad R_\text{trig.} \text{ of type $ B_1^{(1)}$}
    \end{aligned}
\end{equation}
Here $\kappa\in[0, 1]$ plays a role of a deformation parameter and is connected with the $R-$matrix parameter in a certain explicit way. Obviously, $\kappa \rightarrow 0$ produces a round metric and thus the rational $S-$matrix. There is another interesting limit $\kappa\rightarrow 1$, in which the metric becomes flat and $S-$matrix becomes diagonal, which hints to a free fixed point of the RG flow. This turns out to be precisely the case \cite{Fateev:1992tk}: the 1-loop RG flow was shown to reduce to a simple ODE on the only running coupling constant
\begin{equation}
    \frac{d\kappa}{dt} = 1 - \kappa^2 + \text{...} \qquad \Longrightarrow \qquad \kappa(t)_\text{1-loop} = -\tanh t.
\end{equation}
Here by $...$ we mean higher-loop contributions. Parameter $t = \log\frac{\Lambda_0}{\Lambda}$ is the usual RG-time.\footnote{These formulas make sense for $t$ varying from $t_\text{UV} = -\infty$ to $t_\text{0} = 0$, but from the physical perspective one should stop the Ricci flow at some finite time $t_\text{IR}\neq 0$, which corresponds to a dynamically generated mass scale $\Lambda_\text{IR} = \Lambda_0 e^{t_\text{IR}}$, at which the theory becomes strongly coupled and the perturbation theory predictions are not longer reliable. This phenomenon is parallel to the asymptotic freedom of the four-dimensional QCD.}

The similar analysis was then performed for $\OO(4)$ model in \cite{Fateev:1995ht}: it was found that a sigma model action on round 3-dimensional sphere allows for a one-parametric trigonometric deformation resulting in the trigonometric deformation of the corresponding scattering theory:
\begin{equation}
    \begin{aligned}
        ds^2 \sim d\theta^2 + \cos^2\theta\,d\phi^2 + \sin^2\theta\,d\psi^2 &\quad \xrightarrow{\text{trig.def.}}\quad ds^2\sim \frac{d\theta^2 + \cos^2\theta\,d\phi^2}{1 - \kappa^2\sin^2\theta} + \sin^2\theta\, d\psi^2\\
        R_\text{rat.} &\quad\xrightarrow{\text{trig.def.}}\quad R_\text{trig.} \text{ of type $ D_2^{(1)}$}.
    \end{aligned}
\end{equation}
The (generalized) Ricci flow was found to be completely analogous to that of the $\mathrm{O}(3)$ sausage up to some normalization of RG time $t$.

A zero-curvature representation (i.e., classical integrability) for the deformed $\mathrm{O}(3)$ and $\mathrm{O}(4)$ models was not clear at the time. This changed with the celebrated Yang-Baxter (YB) deformation \cite{Klimcik:2008eq, Delduc:2013fga}, which provided a formulation that allowed for a generalization to all $\mathrm{O}(N)$ models with $N > 4$. The action for such a theory reads
\begin{equation}\label{G/H-YB-model}
	S = -\frac{1}{4\lambda^2}\int \tr\left((j_+)^\mathfrak{m} \frac{1}{1-\eta\R_g\,\mathrm P_\mathfrak{m}}(j_-)^\mathfrak{m}\right)d^2x, \qquad (j_\pm)^\mathfrak{m} = \mathrm{P}_\mathfrak{m}(j_\pm),
\end{equation}
where $j_\pm = g^{-1}\partial_\pm g$, $P_\mathfrak{m}$ is an orthogonal projection operator onto coset $\mm$ inside $\gg = \hh\oplus \mm$ and $\RR_g = \Ad_{g}^{-1}\circ\RR\circ\Ad_g$ is a $g$-dependent endomorphism of $\gg = \Lie(\GG)$. YB deformations are one-parameter deformations of PCMs and SSSMs, which are explicitly Lax integrable and correspond to trigonometric $R-$matrices of type $\so_n^{(1)}$ on a dual side.

The modern status is as follows: non-deformed $\OO(N)$ sigma models correspond to rational $R-$matrices, whereas the Yang-Baxter deformed $\OO(N)$ sigma models correspond to the \textit{non-twisted} trigonometric deformations of types $ B_n^{(1)}$ and $ D_n^{(1)}$ (depending on whether $N$ is odd or even) \cite{Litvinov:2018bou,Fateev:2018yos,Alfimov:2020jpy}. A natural question to ask is which sigma models correspond to the \textit{twisted} trigonometric solutions $ D_N^{(2)}\simeq \so_{2N}^{(2)}$. In the present study we fill this gap. Recently one of the authors contributed to defining another deformation of the $\OO(2N)$ model, which was called the dual Yang-Baxter regime \cite{Bychkov:2025ftt}. We claim that it is precisely this model that has the twisted scattering matrix of type $ D_{N}^{(2)}$. For instance, for $\mathrm{O}(4)$ model we claim (in a slightly different choice of coordinates) the following:
\begin{equation}
    \begin{aligned}
        ds^2 \sim d\theta_1^2 + \sin^2\theta_1( d\theta_2^2 + \cos^2\theta_2 d\chi^2) &\quad \xrightarrow{\text{dual regime}}\quad ds^2\sim \frac{d\theta_1^2}{1 - \kappa^2\sin^2\theta_1} +  \sin^2\theta_1\frac{d\theta_2^2 + \cos^2\theta_2\,d\chi^2}{1 - \kappa^2\sin^4\theta_1\sin^2\theta_2}\\
        R_\text{rat.} &\quad\xrightarrow{\text{dual regime}}\quad R_\text{trig., twisted} \text{ of type $ D_2^{(2)}$}.
    \end{aligned}
\end{equation}

Although one-parametric deformations of symmetric space sigma model break the global symmetry group of the action to its maximal torus, it is generally believed that from the integrable systems point of view, the symmetry is not violated but is rather deformed. The usual $G/H$ symmetric space sigma model is naturally associated with the Yangian of $g$ and rational $\mathfrak{g}$-invariant solutions to YBE. Likewise, the deformed $G/H$ coset theories are believed to correspond to quantum groups $U_q(\mathfrak{g})$, which can be viewed as trigonometric deformations of the Yangians, and give rise to the trigonometric $R-$matrices. In this light the YB-deformed $\SO(2N)/\SO(2N-1)$ model corresponds to a quantum group $U_q(\mathfrak{so}_{2N})$. In the present work we relate the corresponding dual regime with the twisted version of $U_q(\mathfrak{so}_{2N})$

The paper is organized as follows. In Section 2 we review the emergence of two different trigonometric solutions of the Yang-Baxter equation, pointing out the apparent coincidence of the trigonometric classical $r$-matrix with the deformation operator in the action of YB-PCM and YB-SSSM. In Section 3 we present our main result: applying the weak-strong duality between sigma models and affine Toda field theories, we compute the tree-level $S-$matrix on the Toda side in the example of $\OO(4)$ model and relate it explicitly to the twisted trigonometric $R-$matrix of type $D_2^{(2)}$. We also calculate the corresponding unitarizing factors (the scattering phase) exactly in Appendix A and use them in the TBA analysis in Appendix B.

\section{Trigonometric deformations of $R-$matrices}
We now briefly sketch how trigonometric deformations appear. Let $\hbar$ be some small real parameter, and consider a family of quantum $R-$matrices depending on it, which we denote by $R_\hbar$. Assuming this dependence to be analytic, we could write $R_\hbar(\theta) = \mathrm{id} + \hbar\, r(\theta) + O(\hbar^2)$. The first order correction $r(\theta)$ is called a classical $r$-matrix. It satisfies the classical Yang-Baxter equation (CYBE)
\begin{equation}
    [r_{12}, r_{13}] + [r_{12}, r_{23}] + [r_{13}, r_{23}] = 0,
\end{equation}
where the dependence on rapidities in $r_{ij}(\theta_i - \theta_j)$ is assumed. Assuming that all quantum $R-$matrices of interest are quasiclassical, i.e. admit such an expansion in powers of $\hbar$, it is natural to first ask for a classification of the classical parts $r(\theta)$ and then try to "quantize" them. In the seminal work \cite{Belavin1998TriangleEA} such a classification result was obtained, and later the quantization problem was solved independently in \cite{Bazhanov:1984gu} and \cite{Jimbo:1985ua}. In order to point out the important features of the trigonometric deformations, which appear at the classical level and persist at the quantum level, we now recall the construction of the trigonometric classical $r$-matrices.

\paragraph{Diagram automorphisms}
Let $\gg$ be a complex finite-dimensional simple Lie algebra. Denote the group of all its automorphisms by $\Aut(\gg)$. Its identity component, $\Aut^0(\gg)$, consists of inner automorphisms and is a normal subgroup of $\Aut(\gg)$. It is generated using the adjoint action (i.e., $x \mapsto \Ad_g(x)$ for some $g \in \GG$ and $x \in \gg = \Lie(\GG)$). The quotient group $\Out(\gg) = \Aut(\gg)/\Aut^0(\gg)$ is called the group of outer automorphisms and is finite for all semisimple Lie algebras. In fact, $\Aut(\gg)/\Aut^0(\gg) \simeq \Aut(\Gamma)$, where $\Gamma$ is the Dynkin diagram of the Lie algebra in question. So $\Out(\gg)$ is trivial for $\gg = B_n, C_n, G_2, F_4, E_7, E_8$; is isomorphic to $\mathbb{Z}_2$ for $\gg = A_n, D_{n\neq 4}, E_6$; and is $S_3$ for $D_4$, a phenomenon sometimes referred to as the triality. The existence of a non-trivial outer automorphism of order 2 for $A_n$, $D_n$, and $E_6$ is precisely what allows for a second trigonometric solution to exist in those cases.

\paragraph{Two trigonometric solutions}
Fix some orthonormal basis $(I_\mu)_{\mu = 1}^{\dim\gg}$ in $\gg$. The Casimir element in $\gg\otimes\gg$ is $t = \sum_\mu I_\mu\otimes I_\mu$. Fix a Dynkin automorphism $\sigma\in\Out(\gg)$ of order $k$. Then $\gg$ as a vector space is decomposed into a direct sum of $\sigma$-eigenspaces: $\gg = \oplus_{j=0}^{k-1}\gg_j$, where $\gg_j = \{x\in\gg:\ \sigma(x) = \omega^j x\}$, $\omega$ is a $k$th root of unity, $\omega^k = 1$. Obviously, $\gg_0$ is a subalgebra. We can use this decomposition to write $t = \sum_{j=0}^{k-1}t_j$, where $t_j\in\gg_j\otimes\gg_j$ is defined by $(\sigma\otimes 1)t_j = \omega^j t_j$. The standard trigonometric solution to CYBE is then \cite{Jimbo:1985ua}
\begin{equation}
	r(x) = r_0 - t_0 + \frac{2}{1-x^k}\sum\limits_{j=0}^{k-1}x^jt_j,
\end{equation}
where $r_0$ is a so-called Drinfeld-Jimbo constant solution in subalgebra
which for $k = 1, 2$\footnote{Which are, in fact, almost all the possible cases. The only unsurprising exception is $D_4 = \so(8)$, which admits a diagram automorphism of order 3 allowing for a third twisted solution $ D_4^{(3)}$.} read
\begin{align}\label{r-classical-Jimbo}
	&k = 1:\qquad r^{(1)}(x) = \frac{1+x}{1-x}\cdot t_0 + r_0\\
	&k = 2:\qquad r^{(2)}(u) = \frac{1+x^2}{1-x^2}(t_0 + t_1) - \frac{1-x}{1+x}t_1 + r_0,
\end{align}
where $r_0$ is usually referred to as Drinfeld-Jimbo $r$-matrix:
\begin{equation}
	r_0 = \sum\limits_{\alpha\in\Delta_0} \mathrm{sgn}(\alpha)\,e_{\alpha}\otimes e_{-\alpha}, \quad \Delta_0 = \{\text{simple roots of $\gg_0$}\}.
\end{equation}
It happens to satisfy the \textit{modified} classical Yang-Baxter equation (mCYBE)
\begin{equation}
	[r_{12}, r_{13}] + [r_{12}, r_{23}] + [r_{13}, r_{23}] = c^2\, \Omega,
\end{equation}
which can be verified by direct computation. Here $\Omega$ is a $\gg$-invariant tensor in $\gg\wedge\gg\wedge\gg$. For simple Lie algebras $\Omega$ is essentially the structure constants tensor: $\Omega_{\mu\nu\lambda} = f_{\mu\nu\lambda}$ (in some orthogonal basis). The usual (non-modified) CYBE is restored in the limit $c^2\rightarrow 0$. 

It is instructive to "raise an index" in $r\in\gg\otimes\gg$ by identifying $\gg\simeq\gg^*$ via the nondegenerate Killing form. Classical $r$-matrix $r\in\gg\wedge\gg$ then becomes $\RR\in\End(\gg)$, mCYBE becomes
\begin{equation}
	[\RR X, \RR Y] + \RR\big([\RR X, Y] + [X, \RR Y]\big) = c^2 [X, Y], \quad \forall X, Y\in\gg,
\end{equation}
and two standard trigonometric solutions become ($x = e^{-2u}$)
\begin{align*}
	&k=1:\qquad \RR^{(1)}(u) = \coth u\cdot\id_\gg + \RR_0,\\
	&k=2:\qquad \RR^{(2)}(u) = \coth2u\cdot \id_\gg - \tanh u\cdot\id_{\gg_0^\perp} + \RR_0.
\end{align*}
In the Cartan-Weyl basis $\{e_\alpha, f_\alpha, h_i\}$ the Drinfeld-Jimbo $r$-matrix $\RR_0$ is
\begin{equation}\label{Drinfeld-Jimbo-sol}
	\RR_0(e_\alpha) = c\,e_{\alpha}, \qquad \RR_0(f_\alpha) = - c\,f_{\alpha}, \qquad \RR_0(h_i) = 0.
\end{equation}
\paragraph{Remark} We note that, quite amazingly, $\RR^{(1)}(u)$ is precisely what enters "the denominator" of the YB-deformed PCM action:
\begin{equation}
	S = -\int\tr\left(j_+\frac{1}{1-\eta\RR_\text{DJ}}j_-\right)d^2x.
\end{equation}
One could pick a closed subgroup $\HH$ inside $\GG$ and gauge out $\hh$-degrees of freedom. This results in the YB-SSSM action \eqref{G/H-YB-model} and the RG-equation for the coupling constant $\eta$:\footnote{The RG equation for YB-PCM (before gauging) is different: $\dot\eta = (1+\eta^2)^2$.}
\begin{equation}
	\dot\eta = h(1+\eta^2) \quad\Longrightarrow\quad \eta(t) = \tanh(h^\vee t),
\end{equation}
where $h^\vee$ is a dual Coxeter number of $\GG$. We see that the initial "denominator" $(\hat{1} + \eta(t)\RR_0)$ is proportional to $\RR^{(1)}(u)$ under the identification of the spectral parameter $u$ with $h^\vee t$. The precise meaning of this coincidence remains unclear (to the best of our knowledge), and the existence of a second integrable deformation given by the twisted solution $\RR^{(2)}(u)$ is an interesting direction of future research.

Treating the solutions \eqref{r-classical-Jimbo} as classical parts of the full quantum $R-$matrices, one could hope to "quantize" all $r(u)$ and find the twisted and untwisted versions of $R(u)$. This is exactly what was done in \cite{Bazhanov:1984gu} and \cite{Jimbo:1985ua} independently for all classical Lie algebras $X_n^{(k)}$ (in fundamental representations of them). We rely on their results heavily in this study.

We once again stress the existence of two different $R-$matrices for $\gg = D_n$. The first one corresponds to a pair $(D_n, \id) =  D_n^{(1)}$ (the non-twisted one), and the other one comes from a nontrivial $\mathbb{Z}_2$-automorphism $(D_n, \sigma) =  D_n^{(2)}$ ($\sigma\neq\id$, the twisted one). The explicit form of both non-twisted and twisted $R-$matrices can be found in \cite{Jimbo:1985ua}, where they are expressed as $R(x, k) = \sum R_{ijkl}(x, k) E_{ij}\otimes E_{kl}$. Here $\{E_{ij}\}$ is the standard basis in $\mathrm{Mat}_{n}(\mathbb{C})$, $x$ is a real multiplicative parameter, which is expected to be related to the rapidities like $x\sim e^\theta$, and $k$ is a parameter of trigonometric deformation, which is expected to be related to the $\lambda$ of \cite{Fateev:1992tk}).

\section{The main result}
To relate a given sigma model with a given quantum $R-$matrix one need to somehow compute the scattering amplitudes in it, at least in some valid approximation. Sigma models are known to be strongly-coupled quantum field theories \cite{Friedan:1980jm}, which prevents one to compute the $S$-matrix via the usual perturbation theory approach, since there is no small parameter available. In the seminal work \cite{Zamolodchikov:1978xm} this difficulty has been overcome by the means of large $N$ expansion with $1/N$ being a small parameter. In this work we take a different route coming from strong-weak type duality. It was first established for the deformed $\OO(3)$-model (sausage) in \cite{Fateev:1992tk}, where it was shown that the duality relates strongly-coupled sigma model regime to a certain integrable perturbation of a sine-Liouville theory. The generalization for $N = 4$ was then obtained in \cite{Fateev:1995ht}, and the generalization for $N > 4$ was established much later in \cite{Fateev:2018yos}. For each $N$ the YB-deformed sigma model was shown to be dual to a certain perturbative integrable field theory called the affine Toda field theory (ATFT). It was shown to describe, after boson-fermion correspondence, the scattering of $\lfloor N/2\rfloor$ pairs of fermions and $\lceil N/2\rceil - 1$ bosons. The scattering matrix was then computed via the bootstrap approach and shown to coincide with the trigonometric solution to YBE of type $\so_N^{(1)}$. 

\subsection{Toda description for the dual regime}
In this section we use the relation between integrable deformed $\OO(N)$ models and ATFTs, which is an example of a weak-strong duality \cite{Fateev:2018yos, Fateev:2019xuq,Litvinov:2018bou}. More explicitly, the sigma model action in the vicinity of the UV fixed point is
\begin{equation}\label{SM-perturb}
    S[X] = \int G_{\mu\nu}(X) \partial X^\mu \bar\partial X^\nu \,d^2x; \qquad G_{\mu\nu}(X) = \delta_{\mu\nu} + \Lambda \sum_{r} A_{\mu\nu}^{(r)} e^{(\boldsymbol{\beta_r},\boldsymbol{X})},
\end{equation}
where $\Lambda$ is some massive parameter (coupling constant), and matrices $A_{\mu\nu}^{(r)}$ always turn out to be of rank one: $A_{\mu\nu}^{(r)} = \alpha_\mu^{(r)}\alpha_\nu^{(r)}$. The structure of vectors $\boldsymbol{\alpha}^{(r)}$ and $\boldsymbol{\beta}^{(r)}$ is almost completely fixed by the integrability and $\mathrm{O}(N)$ symmetry. Namely, the admissible Gram matrix $\Gamma_{ij} = (\boldsymbol{\alpha}_i,\boldsymbol{\alpha}_j)$ is encoded in the following Dynkin-type graph (we only draw the one corresponding to the dual regime):
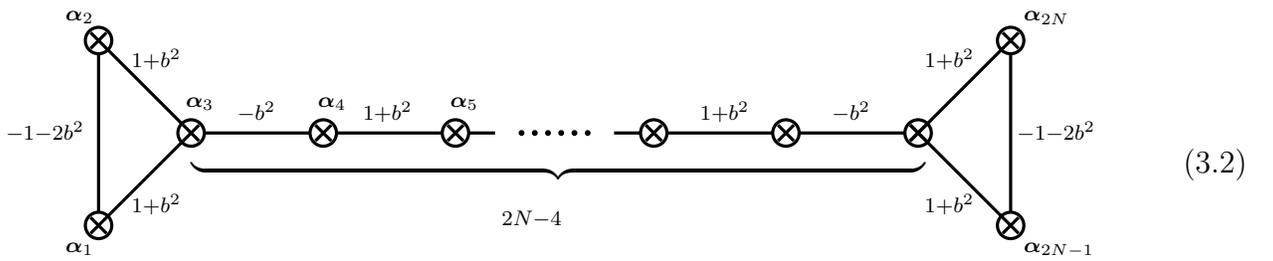
\begin{equation}\label{O(2N)-bantik}
    \begin{picture}(300,60)(260,110)
        \Thicklines
        \unitlength 5pt
        \put(48,32){\circle{2}}
        \put(48,18){\circle{2}}
        \put(54.4,24,4){\line(-1,-1){7}}
        \put(54.4,25,6){\line(-1,1){7}}
        \put(47.4,31.4){\line(1,1){1.2}}
        \put(47.4,18.6){\line(1,-1){1.2}}
        \put(48,19){\line(0,1){12}}
        \put(55,25){\circle{2}}
        \put(54.4,24,4){\line(1,1){1.2}}
        \put(54.4,25,6){\line(1,-1){1.2}}
        \put(66,25){\line(1,0){8}}
        \put(56,25){\line(1,0){8}}
        \put(65,25){\circle{2}}
        \put(64.4,24,4){\line(1,1){1.2}}
        \put(64.4,25,6){\line(1,-1){1.2}}
        \put(75,25){\circle{2}}
        \put(74.4,24,4){\line(1,1){1.2}}
        \put(74.4,25,6){\line(1,-1){1.2}}
        \put(76,25){\line(1,0){2}}
        \put(80,25){\circle{.2}}
        \put(81,25){\circle{.2}}
        \put(82,25){\circle{.2}}
        \put(83,25){\circle{.2}}
        \put(84,25){\circle{.2}}
        \put(85,25){\circle{.2}}
        \put(87,25){\line(1,0){2}}
        \put(90,25){\circle{2}}
        \put(89.4,24,4){\line(1,1){1.2}}
        \put(89.4,25,6){\line(1,-1){1.2}}
        \put(91,25){\line(1,0){8}}
        \put(100,25){\circle{2}}
        \put(99.4,24,4){\line(1,1){1.2}}
        \put(99.4,25,6){\line(1,-1){1.2}}
        \put(101,25){\line(1,0){8}}
        \put(110,25){\circle{2}}
        \put(109.4,24,4){\line(1,1){1.2}}
        \put(109.4,25,6){\line(1,-1){1.2}}
        \put(109.4,24,4){\line(1,1){8.2}}
        \put(109.4,25,6){\line(1,-1){8.2}}
        \put(117,32){\circle{2}}
        \put(117,18){\circle{2}}
        \put(116.4,32.6){\line(1,-1){1.2}}
        \put(116.4,17.4){\line(1,1){1.2}}
        \put(117,19){\line(0,1){12}}
        \put(55,23){$\underbrace{\phantom{aaaaaaaaaaaaaaaaaaaaaaaaaaaaaaaaaaaaaaaaaaaaa}}$}
        \put(78.5,18){$\scriptstyle{2N-4}$}
        \put(50.5,19){$\scriptstyle{1+b^{2}}$}
        \put(50.5,30){$\scriptstyle{1+b^{2}}$}
        \put(41,24.5){$\scriptstyle{-1-2b^{2}}$}
        \put(58.5,26){$\scriptstyle{-b^{2}}$}
        \put(68,26){$\scriptstyle{1+b^{2}}$}
        \put(103.5,26){$\scriptstyle{-b^{2}}$}
        \put(93.5,26){$\scriptstyle{1+b^{2}}$}
        \put(110.5,19){$\scriptstyle{1+b^{2}}$}
        \put(110.5,30){$\scriptstyle{1+b^{2}}$}
        \put(117.5,24.5){$\scriptstyle{-1-2b^{2}}$}
        \put(45.5,16){$\scriptstyle{\boldsymbol{\alpha}_{1}}$}
        \put(45.5,33.5){$\scriptstyle{\boldsymbol{\alpha}_{2}}$}
        \put(54.6,27){$\scriptstyle{\boldsymbol{\alpha}_{3}}$}
        \put(64.6,27){$\scriptstyle{\boldsymbol{\alpha}_{4}}$}
        \put(74.6,27){$\scriptstyle{\boldsymbol{\alpha}_{5}}$}
        \put(118,16){$\scriptstyle{\boldsymbol{\alpha}_{2N-1}}$}
        \put(118,33.5){$\scriptstyle{\boldsymbol{\alpha}_{2N}}$}
    \end{picture}
    \vspace*{1cm}
\end{equation}
Recall that each node $\boldsymbol{\otimes}$ in this diagram encodes a vector $\boldsymbol{\alpha}_i\in\mathbb{C}^{2N-1}$. This $\mathbb{C}^{2N-1}$ is equipped with an $\mathbb{R}$-linear pairing $(-,-)$\footnote{In some orthonormal basis $(\boldsymbol{x},\boldsymbol{y}) = \sum_{k=1}^{2N-1} x_k y_k$ -- no complex conjugation! The "square length" of a given (non-zero) vector $(\boldsymbol{x},\boldsymbol{x})$ can thus be zero, negative or even complex.}, and every $\boldsymbol{\alpha}_i$ is "time-like" with respect to it, that is, $(\boldsymbol{\alpha}_i,\boldsymbol{\alpha}_j) = -1$. Two nodes are connected by a line whenever the corresponding $\boldsymbol{\alpha}$'s pair non-trivially. The value of this pairing $(\boldsymbol{\alpha}_i,\boldsymbol{\alpha}_j)$ depends on a Liouville-like parameter $b$ and is indicated on each line. We enumerate a standard basis of $\mathbb{C}^{2N-1}$ by $\{e_1, \dots , e_{N-1}, E_1, \dots , E_N\}$. For this exact diagram $\boldsymbol{\alpha}$'s can be parametrized as
\begin{equation}\label{alphas-parametrization-general}
    \begin{aligned}
        &\boldsymbol{\alpha}_1 = i\beta e_1 + b E_1, & 
        &\boldsymbol{\alpha}_2 = i\beta e_1 - b E_1,\\
        &\boldsymbol{\alpha}_{2k+1} = -i\beta e_k + b E_{k+1},&
        &\boldsymbol{\alpha}_{2k+2} = i\beta e_{k+1} - bE_{k+1} \quad \text{for}\quad k \in \{1, \dots, N-2\},\\
        &\boldsymbol{\alpha}_{2N-1} = -i\beta e_{N-1} + b E_{N}, &
        &\boldsymbol{\alpha}_{2N} = -i\beta e_{N-1} - b E_{N}.
    \end{aligned}
\end{equation}
It is straightforward to check that the "Gram matrix" of these $\boldsymbol{\alpha}$'s coincides with that of \eqref{O(2N)-bantik}. This is, of course, not the only viable parametrization. 

Next, with each line connecting $\boldsymbol{\alpha}_i$ with $\boldsymbol{\alpha}_j$ we associate the graviton operator
\begin{equation}
    \mathcal{O}_{ij} = (\boldsymbol\alpha_i,\partial_+\boldsymbol\varphi)(\boldsymbol\alpha_i,\partial_-\boldsymbol\varphi) e^{(\boldsymbol\beta_{ij},\boldsymbol{\varphi})}, \qquad\text{where}\quad \boldsymbol\beta_{ij} = \frac{2(\boldsymbol\alpha_i + \boldsymbol\alpha_j)}{(\boldsymbol\alpha_i + \boldsymbol\alpha_j)^2},
\end{equation}
where $\boldsymbol\varphi=(\varphi_1,\dots,\varphi_{2N-1})$ is a multi-component bosonic field. In order for these operators to be IR-relevant, they should have negative anomalous dimension, e.g. $\boldsymbol{\beta}_{ij}$ should be real. That is, the relevant perturbations correspond only to the $(1+b^2)$ edges in our fermionic screening system. The weak-strong duality relates the sigma model
\begin{equation}\label{SM-perturbed-general}
    \mathcal{L}_\textrm{SM} = (\partial\boldsymbol{\varphi},\bar\partial\boldsymbol{\varphi}) + \Lambda \sum\limits_{\begin{smallmatrix}
        \text{real}\\ \text{edges}
    \end{smallmatrix}} (\boldsymbol\alpha_i,\partial\boldsymbol\varphi)(\boldsymbol\alpha_i,\bar\partial\boldsymbol\varphi) e^{(\boldsymbol\beta_{ij},\boldsymbol{\varphi})}
\end{equation}
with the affine Toda field theory
\begin{equation}\label{ATFT-dual-general}
    \mathcal{L}_\textrm{ATFT} = \frac{1}{8\pi}\left(\partial_\mu \varphi\right)^2 + \Lambda\sum_{\textrm{nodes}} e^{(\boldsymbol{\alpha}_k,\boldsymbol{\varphi})}.
\end{equation}

The theory \eqref{SM-perturbed-general} is technically non-renormalizable in a strict sense, requiring introduction of new composite operators in every order of $\Lambda$-perturbation theory. These operators are governed by the RG/Ricci flow
\begin{equation}
    \dot G_{\mu\nu} = -R_{\mu\nu} + 2\,\nabla_\mu\nabla_\nu\Phi.
\end{equation}
Assuming its solution to be analytic in $\Lambda$, it is possible to find the full sigma model metric $G_{\mu\nu}$\footnote{In fact, the T-duality in all isometric directions is required. It produces a purely-imaginary $B-$field.}. For this exact fermionic screening system the metric and the $B-$field coincide with that of the dual regime\footnote{The usual (untwisted) regime is given by the same bowtie, with different choice of coupling $b^2 \mapsto -1-b^2$ and perturbing set of operators.} of YB-deformed sigma model after some change of coordinates \cite{Bychkov:2025ftt}. In this way, the duality between the (dual) YB-deformed spheres and ATFTs is established. The transition to the dual ATFT, which is inherently perturbative in $\Lambda$, provides for another way (other than $1/N$ expansion) to compute scattering amplitudes in a given deformed sigma model via the conventional Feynman rules.

In this work, we apply this technique to the dual regime of the YB-deformed sphere sigma model. We stress once again that this dual regime only exists for odd-dimensional spheres $\SS^{2N-1} = \frac{\OO(2N)}{\OO(2N-1)}$, since the $\OO(2N) \simeq D_N$ of the "numerator" allows for an outer $\mathbb{Z}_2$-automorphism, while $\OO(2N+1) \simeq B_N$ does not (see \cite{Bychkov:2025ftt} for the details). 

\subsection{Dual twisted Toda Lagrangians}
We start with the ATFT Lagrangian \eqref{ATFT-dual-general}. It is convenient to label the fields according to the parametrization \eqref{alphas-parametrization-general}: 
\begin{equation}
    (E_k,\boldsymbol\varphi) = \Phi_k, \qquad (e_k,\boldsymbol\varphi) = \phi_k.
\end{equation}
We apply the boson-fermion correspondence \cite{Coleman:1975, Mandelstan:1975} to $\phi_k$: 
\begin{equation}
    \frac{1}{8\pi}\left(\partial_\mu \phi_k\right)^2 \to i\overline{\psi}_k\gamma^\nu\partial_\nu\psi_k+\frac{\pi b^2}{2(1+b^2)}\left(\overline{\psi}_k\gamma^\nu\psi_k\right)^2,\qquad e^{\pm i\beta\phi_k}\to \overline{\psi}_k\frac{1\pm\gamma_5}{2}\psi_k=\overline{\psi}_k\gamma_\pm\psi_k.
\end{equation}
With this identification, the Lagrangian \eqref{ATFT-dual-general} transforms into its Toda-Thirring form
\begin{multline}
    \mathcal{L}=\frac{1}{8\pi}\sum_{k=1}^N\left(\partial_\mu\Phi_k\right)^2 + \sum_{k=1}^{N-1}\left[i\overline{\psi}_k\hat{\partial}\psi_k+\frac{\pi b^2}{2(1+b^2)}\left(\overline{\psi}_k\gamma^\nu\psi_k\right)^2\right]-\\
    -m\left\{ \overline{\psi}_1\gamma_+\psi_1\cosh b\Phi_1 +\sum_{k=1}^{N-2}\left(e^{b\Phi_{k+1}}\overline{\psi}_k\gamma_-\psi_k+e^{-b\Phi_{k+1}}\overline{\psi}_{k+1}\gamma_+\psi_{k+1}\right)+\overline{\psi}_{N-1}\gamma_-\psi_{N-1} \cosh b\Phi_N  \right\} -\\
    -\frac{m^2}{4\pi b^2}\left\{e^{b\Phi_2}\cosh b\Phi_1 +\sum_{k=2}^{N-2}e^{b(\Phi_{k+1}-\Phi_{k})} + e^{-b\Phi_{N-1}}\cosh b\Phi_N - (N-1)\right\}.
\end{multline}
Next, we expand this theory to the first nontrivial order in $b^2$, diagonalize the mass matrix and compute the tree-level scattering amplitudes. We claim that the resulting scattering matrix coincides with the first-order approximation of $D_N^{(2)}$ trigonometric $R-$matrix modulo unitarizing factor $F(\theta)$.


\subsection{Explicit example: $\mathrm{O}(4)$ model} 
We now implement the procedure explained above to the simplest non-trivial example of the dual regime of the YB-deformed $\OO(4)$ model. We claim that the $S-$matrix of the corresponding ATFT coincides with quantum $R-$matrix of type $D^{(2)}_{2}$. Since the solution to YBE need not be a unitary and crossing-invariant, we introduce a unitarizing function $F(\theta)$: 
\begin{gather}
    S_{i_1.j_1}^{i_2,j_2}(\theta) = F(\theta) R_{i_1.j_1}^{i_2,j_2}(\theta),
    \label{eq:RScorrespondence}
\end{gather}
which we will find explicitly later.

Following our construction of the dual regime, we consider the fermionic root system
\begin{equation}\label{O(4)-bantik-dual}
\begin{picture}(300,60)(140,110)
    \Thicklines
    \unitlength 5pt
    \put(48,32){\circle{2}}
    \put(48,18){\circle{2}}
    \put(62.7,32.7){\line(-1,-1){15.4}}
    \put(49,32.1){\line(1,0){12}}
    \put(49,18){\line(1,0){12}}
    \put(62.8,17.2){\line(-1,1){15.4}}
    \put(47.4,31.4){\line(1,1){1.2}}
    \put(47.4,18.6){\line(1,-1){1.2}}
    \put(48,19){\line(0,1){12}}
    \put(62,32){\circle{2}}
    \put(62,18){\circle{2}}
    \put(61.4,32.6){\line(1,-1){1.2}}
    \put(61.4,17.4){\line(1,1){1.2}}
    \put(62,19){\line(0,1){12}}
    \put(50.5,19){$\scriptstyle{1+b^{2}}$}
    \put(50.5,30){$\scriptstyle{1+b^{2}}$}
    \put(53.5,33){$\scriptstyle{1+b^{2}}$}
    \put(53.5,16){$\scriptstyle{1+b^{2}}$}
    \put(45.5,16){$\scriptstyle{\boldsymbol{\alpha}_{1}}$}
    \put(45.5,33.5){$\scriptstyle{\boldsymbol{\alpha}_{2}}$}
    \put(62.5,16){$\scriptstyle{\boldsymbol{\alpha}_{3}}$}
    \put(62.5,33.5){$\scriptstyle{\boldsymbol{\alpha}_{4}}$}
    \put(56,19){$\scriptstyle{1+b^{2}}$}
    \put(56,30){$\scriptstyle{1+b^{2}}$}
    \put(41,24.5){$\scriptstyle{-1-2b^{2}}$}
    \put(62.5,24.5){$\scriptstyle{-1-2b^{2}}$}
  \end{picture}
  \vspace*{1cm}
\end{equation}
and parametrize $\boldsymbol{\alpha}$'s in the following way:
\begin{align}
    \boldsymbol\alpha_1 &= -b E_1+i\beta e_1, \qquad \boldsymbol\alpha_2 = bE_1+i\beta e_1,\\ 
    \boldsymbol\alpha_3 &= b E_2-i\beta e_1,\qquad\ \ \,  \boldsymbol\alpha_4 = -bE_2-i\beta e_1.
\end{align}
The corresponding ATFT Lagrangian \eqref{ATFT-dual-general} is
\begin{multline}
    \mathcal{L} = \frac{1}{8\pi}\sum_{k=1}^2 (\partial_\mu \phi)^2 + i\overline{\psi}\hat{\partial}\psi + \frac{\pi b^2}{2(1+b^2)}(\overline{\psi}\gamma^\mu\psi)^2 -\\
    -m\left[\cosh b\phi_1 \overline{\psi}\gamma_+\psi + \cosh b\phi_2 \overline{\psi}\gamma_-\psi \right] -\frac{m^2}{4\pi b^2}(\cosh b\phi_1 \cosh b\phi_2 - 1)
\end{multline}
Expanding it to the first nontrivial order in $b^2 \to 0$, we obtain
\begin{multline}
    \mathcal{L} = \frac{1}{8\pi}\sum_{k=1}^2\left((\partial_\mu\phi_k)^2 - m^2\phi_k^2\right) + \overline{\psi}(i\hat{\partial} + \hat{A} -m)\psi + \frac{1+b^2}{2\pi b^2}A^2 -  \\ - \frac{mb^2}{2!}\left[\phi_1^2\overline{\psi}\gamma_+\psi + \phi_2^2\overline{\psi}\gamma_-\psi\right]-\frac{m^2 b^2}{4\pi\cdot 4!}(\phi_1^4+6\phi_1^2\phi_2^2+\phi_2^4)+ O(b^4), 
\end{multline}
where we introduced an auxiliary field $A_\mu = \frac{(\sqrt{2}-1) \pi b^2}{1+b^2}\, \overline{\psi}\gamma_\mu\psi$ for the sake of convenience. All fundamental degrees of freedom in this theory constitute a single multiplet of fields of mass $m$. 

A straighforward computation yields a tree-order two-particle $S-$matrix, which we write down in the basis $\{\psi,\phi_1,\phi_2,\overline{\psi}_1\}=\{+,1,2,-\}$
\begin{equation}
    S_\text{pert}^{(1)}(\theta)=\left(
    \begin{smallmatrix}
        ++ & +1 & +2 & +- & 1+ & 11 & 12 & 1- & 2+ & 21 & 22 & 2- & -+ & -1 & -2 & -- \\ \hline \\
        a_\theta & 0 & 0 & 0 & 0 & 0 & 0 & 0 & 0 & 0 & 0 & 0 & 0 & 0 & 0 & 0 \\ 
        0 & b_\theta & 0 & 0 & -i f_{-\theta} & 0 & 0 & 0 & 0 & 0 & 0 & 0 & 0 & 0 & 0 & 0 \\ 
        0 & 0 & b_\theta & 0 & 0 & 0 & 0 & 0 & if_\theta & 0 & 0 & 0 & 0 & 0 & 0 & 0 \\ 
        0 & 0 & 0 & a_{i\pi-\theta} & 0 & f_{-\theta} & 0 & 0 & 0 & 0 & f_\theta & 0 & 2b_\theta & 0 & 0 & 0 \\ 
        0 & i f_\theta & 0 & 0 & b_\theta & 0 & 0 & 0 & 0 & 0 & 0 & 0 & 0 & 0 & 0 & 0 \\ 
        0 & 0 & 0 & f_\theta & 0 & b_\theta & 0 & 0 & 0 & 0 & b_{i\pi-\theta} & 0 & if_{i\pi-\theta} & 0 & 0 & 0 \\ 
        0 & 0 & 0 & 0 & 0 & 0 & b_\theta & 0 & 0 & b_\theta & 0 & 0 & 0 & 0 & 0 & 0 \\ 
        0 & 0 & 0 & 0 & 0 & 0 & 0 & b_\theta & 0 & 0 & 0 & 0 & 0 & f_{i\pi-\theta} & 0 & 0 \\ 
        0 & 0 & -if_{-\theta} & 0 & 0 & 0 & 0 & 0 & b_\theta & 0 & 0 & 0 & 0 & 0 & 0 & 0 \\ 
        0 & 0 & 0 & 0 & 0 & 0 & b_\theta & 0 & 0 & b_\theta & 0 & 0 & 0 & 0 & 0 & 0 \\ 
        0 & 0 & 0 & f_{-\theta} & 0 & b_\theta & 0 & 0 & 0 & 0 & b_\theta & 0 & -if_{\theta-i\pi} & 0 & 0 & 0 \\ 
        0 & 0 & 0 & 0 & 0 & 0 & 0 & 0 & 0 & 0 & 0 & b_{i\pi-\theta} & 0 & 0 & f_{\theta-i\pi} & 0 \\ 
        0 & 0 & 0 & 2b_\theta & 0 & -i f_{\theta-i\pi} & 0 & 0 & 0 & 0 & if_{i\pi-\theta} & 0 & a_{i \pi-\theta} & 0 & 0 & 0 \\ 
        0 & 0 & 0 & 0 & 0 & 0 & 0 & if_\theta & 0 & 0 & 0 & 0 & 0 & b_{i\pi-\theta} & 0 & 0 \\ 
        0 & 0 & 0 & 0 & 0 & 0 & 0 & 0 & 0 & 0 & 0 & -i f_{-\theta} & 0 & 0 & b_{i\pi-\theta} & 0 \\ 
        0 & 0 & 0 & 0 & 0 & 0 & 0 & 0 & 0 & 0 & 0 & 0 & 0 & 0 & 0 & a_\theta
    \end{smallmatrix}.
    \right)
    \label{eq:PertSMatrix}
\end{equation}
with entries 
\begin{equation}
    a_\theta = -i\pi\tanh\frac{\theta}{2},\quad b_\theta = -\frac{\pi i}{\sinh \theta}, \quad  f_\theta = -\pi\frac{e^{\theta/2}}{\sinh\theta}.
\end{equation}
We note that this $S-$matrix is manifestly $\mathbf{C}$- and $\mathbf{PT}$-invariant. The $\mathbf{P}$ and $\mathbf{T}$ symmetries are broken, which alignes with the classical treatment of these sigma models: the Yang-Baxter deformation forces the introduction of a nontrivial purely imaginary $B-$field, which certainly spoils hermicity of the corresponding Hamiltonian as well as the symmetry under the change of orientation on the world-sheet ($\mathbf{P}$ and $\mathbf{T}$, but not $\mathbf{PT}$).

Now we match this result to the quantum $R-$matrix of type $D_2^{(2)}$, which we borrow from \cite{Jimbo:1985ua}. The explicit form of it is too large of a matrix to be presented here, so we do not present it here. Instead, we stress that it depends on two variables: $x$ and $k$. The former is the spectral parameter, we expect it to be related with the rapidity $\theta$ in a trigonometric manner; likewise, the deformation parameter $k$ is expected to be a function of $\lambda$, which in turn depends on the coupling constant $b^2$ (see \cite{Alfimov:2020jpy} for the details). We assume these dependencies to be of the simplest form
\begin{equation}
    x=\exp\left(\alpha\lambda\theta\right),\qquad k=\exp\left(i\beta\lambda\right)
\end{equation}
with constant $\alpha$ and $\beta$, which are yet to be determined. This $R-$matrix has, in fact, two free (trivial scattering) points: at $k = 1$ (or $\lambda = 0$) and at $k = -1$. The former fixed point corresponds to a non-deformed limit, whereas the latter accounts for a UV theory. We will also assume that, in agreement with the previous considerations, $k = -1$ fixed point corresponds to $\lambda = 1/2$. In order to relate $S(\theta)_\text{pert}^{(1)}$ to $F(\theta,\lambda)R(\theta,\lambda)$, we expand both the $R-$matrix and the unitarizing function $F(\theta)$ in the vicinity of the UV fixed point $\lambda\rightarrow1/2$ to the first order:
\begin{equation}
    F(\theta, \lambda)=F^{(0)}(\theta)+F^{(1)}(\theta)(\lambda-1/2)+O(\lambda-1/2)^2.
\end{equation}
This $FR$ matches with the $S-$matrix up to some linear transformation in the space of fields, namely, the of rotation of bosons $\{\phi_1,\phi_2\}\mapsto\{\frac{-1}{\sqrt{2}}(\phi_1+i\phi_2),\frac{-1}{\sqrt{2}}(\phi_1-i\phi_2)\}$ and the appropriate rescaling of fermions. The combined transformation results in conjugation by the matrix
\begin{equation}\label{M(theta)}
    M(\theta) = \left(
    \begin{matrix}
        e^{-\theta \lambda} & 0 & 0 & 0\\
        0 & -\frac{1}{\sqrt{2}} & -\frac{i}{\sqrt{2}} & 0\\
        0 & -\frac{1}{\sqrt{2}} & \frac{i}{\sqrt{2}} & 0\\
        0 & 0 & 0 & e^{\theta \lambda}
    \end{matrix}
    \right).
\end{equation}
That is, instead of $R(\theta,\lambda)$ we consider
\begin{equation}\label{Rtilde-conj}
    \Tilde{R}(\theta,\lambda) = \big(M(\theta_1)\otimes M(\theta_2)\big)R(\theta_1 - \theta_2,\lambda)\left(M(\theta_1)\otimes M(\theta_2)\right)^{-1},
\end{equation}
or, more explicitly,
\begin{equation}
    \Tilde{R}^{i_1i_2}_{j_1j_2} (\theta,\lambda) = F(\theta, \lambda)\left[M^{-1}(\theta_2)\right]^{p_1}_{j_1}\left[M^{-1}(\theta_1)\right]^{p_2}_{j_2}R^{q_1q_2}_{p_1p_2}(\theta_1 - \theta_2, \lambda)\left[M(\theta_1)\right]^{i_1}_{q_1}\left[M(\theta_2)\right]^{i_2}_{q_2}.
\end{equation}

Now we determine the values of $\alpha$ and $\beta$ by comparing the first two orders in $(\lambda - \frac{1}{2})$-expansion of $\Tilde{R}$ with that of the tree-level $S-$matrix (\ref{eq:PertSMatrix}). Treating the equation $F(\theta)\Tilde{R}(\theta,\lambda) = S_\text{pert}(\theta)$ perturbatively in the 0th and 1st order in $(\lambda - \frac{1}{2})$, we arrive at 
\begin{align}
    S_\text{pert}^{(0)}(\theta) &= F^{(0)}(\theta) \Tilde{R}^{(0)}(\alpha,\beta,\theta),\\
    S_\text{pert}^{(1)}(\theta) &= F^{(0)}(\theta) \Tilde{R}^{(1)}(\alpha,\beta,\theta) + F^{(1)}(\theta) \Tilde{R}^{(0)}(\alpha,\beta,\theta)(\lambda-1/2).
\end{align}
From this overdetermined system of algebraic equations we extract the four unknown functions: $F^{(0)}$, $F^{(1)}$, $\alpha$, $\beta$:
\begin{equation}
    \alpha = 2,\qquad \beta = 2\pi,\qquad F^{(0)}(\theta) = -\left(\frac{e^{-\theta}}{2\sinh\theta}\right)^2,\qquad F^{(1)}(\theta) = \frac{8 e^{2 \theta } \theta +2 i \pi  \left(e^{\theta }-3\right) \left(e^{\theta }+1\right)}{\left(e^{2 \theta }-1\right)^3}
\end{equation}
That fixes the parametrization of $x$ and $k$:
\begin{gather}
    x = \exp(2\lambda\theta),\qquad k = \exp(2i\pi \lambda).
\end{gather}

Now, using this parametrization and assuming the correspondence (\ref{eq:RScorrespondence}) to hold, we determine the unitarizing function $F(\theta)$ exactly. The unitarity and crossing-invariance conditions are
\begin{equation}
    S_{12}(\theta)S_{21}(-\theta)=1,\qquad S_{12}(\theta)=C_1^{-1} S_{2\tilde{1}}(i\pi-\theta)C_1,
\end{equation}
where $C_1$ is charge conjugation matrix acting in the first space. These conditions impose functional equations on $F(\theta)$:
\begin{align}\label{eq:UnitFunction}
    &F(\theta)F(-\theta)=1,\\
    &F(i\pi - \theta)=F(\theta)\frac{k^2(x^2-1)^2}{(k^2-x^2)^2} =F(\theta)\frac{\sinh^2(2\theta\lambda)}{\sinh^2(2\lambda(\theta-i\pi))}
\end{align}
These formulas are derived for dual YB-deformed $\OO(4)$ model. For general $\OO(2n)$ we conjecture the following relation, which can be checked by analogous arguments:
\begin{equation}\label{F(ipi-theta)=F(theta)}
    F(i\pi - \theta) = F(\theta)\frac{\sinh \left(n \lambda  \left(\theta -\frac{i \pi  (n-2)}{n}\right)\right) \sinh \left(n\lambda \theta\right)}{\sinh \left(n \lambda \left(\theta -\frac{2 i \pi }{n}\right)\right) \sinh \left(n \lambda (\theta - i \pi )\right)}.
\end{equation}
The calculation producing the minimal solution to these equations is presented in Appendix A. The answer for general $\OO(2n)$ is
\begin{equation}
    \log F(\theta) = i\int_{-\infty}^\infty \frac{d\omega}{\omega} \frac{\cosh\frac{\pi\omega (n-2)}{2n}\sinh\frac{\pi\omega}{n}(\frac{1}{2\lambda}-1)}{\cosh\frac{\pi\omega}{2}\sinh\frac{\pi\omega}{2n\lambda}}\sin\theta\omega.
\end{equation}
It happens to coincide with the scattering phase in the usual YB-regime presented in \cite{Fateev:2018yos}. We note that in the limit $\lambda\to 0$ one recovers the scattering matrix of the non-deformed $\OO(2n)$ model, corresponding to the rational solution to quantum YBE.

We also performed similar calculations for dual-YB-$\OO(6)$ model. The results are quite similar to that of the simplest $\OO(4)$ case, but are, perhaps, too long to be presented here ($S-$matrix is $36\times 36$). Instead, we only sketch what wee expect to happen for a general $\OO(2n)$ situation. The dual-YB-$\OO(2n)$ sigma model possesses a total of $n-1$ isometries, which are fermionized on the Toda side \eqref{ATFT-dual-general} $\phi_k \leftrightarrow(\psi_k, \bar\psi_k)$, and a total of $n$ bosonic fields $\Phi_k$. We are interested in the scattering of the lightest multiplet of fundamental particles. After expansion to the linear order in $b^2$ and mass matrix diagonalization, we are left with a multiplet of $2\times(n-1)$ fermions plus $2$ bosons, a total of $2n$ lightest fields. Then one has to calculate the tree-level $S-$matrix in the basis $(\psi_1, ..., \psi_{n-1}, \Phi_1, \Phi_2, \bar\psi_{n-1},...,\bar\psi_1)$ and relate it to the twisted $R-$matrix of type $D_n^{(2)}$ in the fundamental representation (both of them are $(2n)^2\times(2n)^2$ matrices). Again, $S_\text{tree}$ and $R(\lambda,\theta)$ are equal only in the first order in $\lambda$, up to a multiplicative unitarizing factor $F(\theta)$, which we presented above, and a up to a conjugation by $M(\theta_1)\otimes M(\theta_2)$ analogous to \eqref{M(theta)}, \eqref{Rtilde-conj}. We expect $M(\theta)$ to be diagonal with entries $e^{\pm k\theta\lambda}$ expect for a $2\times2$ block of "45-degree rotation" in the center.

The scattering phase $F(\theta)$ can be further exploited to derive physical properties of this field system, such as the ground energy density and mass renormalization via the technique of Thermodynamic Bethe Ansatz (TBA). Since $F(\theta)$ in our dual YB regime does not differ from that of the usual regime, the answers in the twisted and untwisted regimes are not expected to be terribly different. Nevertheless, we present the corresponding calculations in Appendix B.

\section{Concluding remarks}
In this work we studied the quantum properties of the newly introduced dual (or twisted) regime in Yang-Baxter deformed sphere sigma models \cite{Bychkov:2025ftt}. In particular, we computed all tree-level scattering amplitudes in the dual-YB-$\OO(4)$ model, the first nontrivial example of dual regimes, and explicitly related the corresponding scattering matrix with the twisted trigonometric solution of type $D_2^{(2)}$\footnote{We also checked the analogous conjecture for some amplitudes in dual-YB-$\OO(6)$ and dual-YB-$\OO(8)$ models, but the calculations are too bulky to be presented in this short text.}. This motivates the general conjecture: the scattering in dual-YB-$\OO(2N)$ model is described by a twisted trigonometric $R-$matrix of type $D_N^{(2)}$. The important questions for future study include:
\begin{itemize}
    \item The emergence of a second, twisted, trigonometric solution is related to the nontrivial outer automorphism $\sigma\in\Out(\gg)$. For almost all simply-laced classical Lie algebras it has order 2. The only exception is $D_4 \simeq \so_8$, for which the group of outer automorphisms is isomorphic to $S_3$ and has an element of order 3. It is thus very natural to ask for a $\mathbb Z_3$-twisted ATFT, which would describe this kind of scattering, and ultimately for a sigma model description.
    \item The next important point is related to the content of the Remark after \eqref{Drinfeld-Jimbo-sol}. The Lagrangian description of the dual regimes is unknown. One could argue that the general recipe is to start from a general one-parametric deformation of PCM action
    \begin{equation}
        \mathcal{A} = -\int\mathrm{tr}\left(J_+\,J_-\right)d^2x \quad \longmapsto \quad \mathcal{A}(\lambda) = -\int\mathrm{tr}\left(J_+\,\Omega(\lambda)\,J_-\right)d^2x \quad \xmapsto{J\mapsto J - A}\quad \text{coset model}
    \end{equation}
    and assume $\Omega^{-1}(\lambda)$ to be the trigonometric parameter-dependent classical $r$-matrix. For the untwisted $r-$matrix we get the usual Yang-Baxter deformation: $r(u) = 1 + \tanh u\cdot\mathcal{R}$ ($u$ plays the role of RG time). The question is what one would get for the twisted version $r(u) = 1 - \tanh u\tanh 2u\cdot 1_\mathfrak{m} + \tanh 2u\cdot\mathcal{R}$. Is this sigma model classically integrable? If so, what does the Lax connection look like? Is it related to our dual regime?
    \item It is natural to ask for a similar construction (duality with ATFT, dual regimes, Lax representation, etc) for other series of symmetric cosets. The starting point could be (one of) the Grassmann manifolds:
    \begin{itemize}
        \item Real Grassmannians $\SO(p+q)/\SO(p)\times\SO(q)$
        \item Complex Grassmannians $\mathrm{SU}(p+q)/\mathrm{S}(\mathrm{U}(p)\times\mathrm{U}(q))$
        \item Real Lagrangian Grassmannians $\mathrm{SU}(n)/\SO(n)$
        \item Complex Lagrangian Grassmannians $\mathrm{USp}(2n)/\mathrm{U}(n)$
    \end{itemize}
    The UV fixed point CFT, relevant perturbation, conserved charges, etc.\ are expected to be quite different from that of $\SO(N)/\SO(N-1)$ cosets. For instance, we found the UV fixed point CFT of the YB-deformed $\mathrm{SU}(3)/\mathrm{SO}(3)$ sigma model to be no longer a free theory, but rather a $\mathrm{SL}(2, \mathbb{R})\times\mathrm{U}(1)^2$ WZW. We expect the fixed point of the YB-deformed $\mathrm{SU}(n)/\SO(n)$ to be $\mathrm{SO}(1, n-1)\times \mathrm{U}(1)^{n-1}$ WZW model.
\end{itemize}

\section*{Acknowledgments} We are thankful to Anton Pribytok, Mikhail Alfimov, Dmitry Bykov and Ben Hoare for helpful discussions. We are especially indebted to Alexey Litvinov for the collaboration on the precursor of this paper, his continuous support and kind scientific advisory.

\appendix\section{Scattering phase $F(\theta)$}
We need to find the so-called "minimal solution" to \eqref{F(ipi-theta)=F(theta)}. There are various ways to do this; we will use the most straightforward one mostly following \cite{Bombardelli}.

Using the representation of hyperbolic functions via gamma functions:
\begin{align}
    \sinh \pi z &= \frac{\pi}{\Gamma \left(1+i z\right)\Gamma\left(-i z\right)}\\
    \cosh \pi z &= \frac{\pi}{\Gamma\left(\frac{1}{2}+i z\right)\Gamma\left(\frac{1}{2}-iz\right)}
\end{align}
we rewrite the second equation in (\ref{eq:UnitFunction}) as
\begin{gather}
    F(i\pi - \theta)=F(\theta)\frac{\Gamma\left(1+i\frac{n\lambda}{\pi}\left(\theta-\frac{2i\pi}{n}\right)\right)\Gamma\left(-i\frac{n\lambda}{\pi}\left(\theta-\frac{2i\pi}{n}\right)\right)\Gamma\left(1+i\frac{n\lambda}{\pi}\left(\theta-i\pi\right)\right)\Gamma\left(-i\frac{n\lambda}{\pi}\left(\theta-i\pi\right)\right)}{\Gamma\left(1+i\frac{n\lambda}{\pi}\left(\theta-\frac{(n-2)i\pi}{n}\right)\right)\Gamma\left(-i\frac{n\lambda}{\pi}\left(\theta-\frac{(n-2)i\pi}{n}\right)\right)\Gamma\left(1+i\frac{n\lambda}{\pi}\theta\right)\Gamma\left(-i\frac{n\lambda}{\pi}\theta\right)} = \\ = F(\theta)\frac{\Gamma\left(1+2\lambda + i\frac{n\lambda \theta}{\pi}\right)\Gamma\left(-2\lambda - i\frac{n\lambda \theta}{\pi}\right)\Gamma\left(1+n\lambda+i\frac{n\lambda\theta}{\pi}\right)\Gamma\left(-n\lambda-i\frac{n\lambda\theta}{\pi}\right)}{\Gamma\left(1+(n-2)\lambda + i\frac{n\lambda\theta}{\pi}\right)\Gamma\left(-(n-2)\lambda - i\frac{n\lambda\theta}{\pi}\right)\Gamma\left(1+i\frac{n\lambda\theta}{\pi}\right)\Gamma\left(-i\frac{n\lambda\theta}{\pi}\right)}.
    \label{eq:shiftF}
\end{gather}
We start with the most simple ansatz
\begin{equation}
    F(\theta) = F_1(\theta) = F_2(\theta)\frac{\Gamma\left(1+(n-2)\lambda+i\frac{n\lambda\theta}{\pi}\right)\Gamma\left(1+i\frac{n\lambda\theta}{\pi}\right)}{\Gamma\left(2\lambda+i\frac{n\lambda\theta}{\pi}\right)\Gamma(n\lambda+i\frac{n\lambda\theta}{\pi})},
\end{equation}
where the function $F_2(\theta)$ satisfies $F_2(\theta)=F_2(i\pi-\theta)$. Indeed, writing
\begin{equation}
    F_1(i\pi-\theta)=F_2(i\pi-\theta)\frac{\Gamma\left(1-2\lambda-i\frac{n\lambda\theta}{\pi}\right)\Gamma\left(1-n\lambda-i\frac{n\lambda\theta}{\pi}\right)}{\Gamma\left(-(n-2)\lambda-i\frac{n\lambda\theta}{\pi}\right)\Gamma(-i\frac{n\lambda\theta}{\pi})},
\end{equation}
we arrive at
\begin{equation}
    \frac{F_1(i\pi-\theta)}{F_1(\theta)} =\frac{\Gamma\left(2\lambda + i\frac{n\lambda \theta}{\pi}\right)\Gamma\left(1-2\lambda - i\frac{n\lambda \theta}{\pi}\right)\Gamma\left(n\lambda+i\frac{n\lambda\theta}{\pi}\right)\Gamma\left(1-n\lambda-i\frac{n\lambda\theta}{\pi}\right)}{\Gamma\left(1+(n-2)\lambda + i\frac{n\lambda\theta}{\pi}\right)\Gamma\left(-(n-2)\lambda - i\frac{n\lambda\theta}{\pi}\right)\Gamma\left(1+i\frac{n\lambda\theta}{\pi}\right)\Gamma\left(-i\frac{n\lambda\theta}{\pi}\right)},
\end{equation}
which, after using $\Gamma(1+z) = z\Gamma(z)$ in the numerator, matches with the equation (\ref{eq:shiftF}) exactly. 

The unitarity condition $F(\theta)F(-\theta) = 1$ for $F_1(\theta)$ (the first equation of \ref{eq:UnitFunction}) is inherited by $F_2(\theta)$ as well. Indeed,
\begin{equation}
    F_2(\theta)F_2(-\theta)\frac{\Gamma\left(1+(n-2)\lambda+i\frac{n\lambda\theta}{\pi}\right)\Gamma\left(1+i\frac{n\lambda\theta}{\pi}\right)}{\Gamma\left(2\lambda+i\frac{n\lambda\theta}{\pi}\right)\Gamma(n\lambda+i\frac{n\lambda\theta}{\pi})}\frac{\Gamma\left(1+(n-2)\lambda-i\frac{n\lambda\theta}{\pi}\right)\Gamma\left(1-i\frac{n\lambda\theta}{\pi}\right)}{\Gamma\left(2\lambda-i\frac{n\lambda\theta}{\pi}\right)\Gamma(n\lambda-i\frac{n\lambda\theta}{\pi})} \equiv 1.
\end{equation}
Once again, using the most simple ansatz, we write
\begin{equation}
    F_2(\theta) = F_3(\theta)\frac{\Gamma\left(2\lambda-i\frac{n\lambda\theta}{\pi}\right)\Gamma(n\lambda-i\frac{n\lambda\theta}{\pi})}{\Gamma\left(1+(n-2)\lambda-i\frac{n\lambda\theta}{\pi}\right)\Gamma\left(1-i\frac{n\lambda\theta}{\pi}\right)}.
\end{equation}
Using unitarity $F_2(\theta)=F_2(i\pi-\theta)$, we obtain an equation for $F_3(\theta)$. This process can be continued indefinetely. It has a period of 4, which results in the following solution:
\begin{multline}
    F(\theta)=-\prod_{k=0}^{\infty}\frac{\Gamma\left(1+[(2k+2)n-2]\lambda-i\frac{n\lambda\theta}{\pi}\right)}{\Gamma\left(1+[(2k+2)n-2]\lambda+i\frac{n\lambda\theta}{\pi}\right)}\frac{\Gamma\left(1+(2k+1)n\lambda-i\frac{n\lambda\theta}{\pi}\right)}{\Gamma\left(1+(2k+1)n\lambda+i\frac{n\lambda\theta}{\pi}\right)}\times\\\times\frac{\Gamma\left([(2k+1)n+2]\lambda+i\frac{n\lambda\theta}{\pi}\right)}{\Gamma\left([(2k+1)n+2]\lambda-i\frac{n\lambda\theta}{\pi}\right)}\frac{\Gamma\left((2k+2)n\lambda+i\frac{n\lambda\theta}{\pi}\right)}{\Gamma\left((2k+2)n\lambda-i\frac{n\lambda\theta}{\pi}\right)}\frac{\Gamma\left([2kn+2]\lambda-i\frac{n\lambda\theta}{\pi}\right)}{\Gamma\left([2kn+2]\lambda+i\frac{n\lambda\theta}{\pi}\right)}\nonumber\times\\\times\frac{\Gamma\left((2k+1)n\lambda-i\frac{n\lambda\theta}{\pi}\right)}{\Gamma\left((2k+1)n\lambda+i\frac{n\lambda\theta}{\pi}\right)}\frac{\Gamma\left(1+[(2k+1)n-2]\lambda+i\frac{n\lambda\theta}{\pi}\right)}{\Gamma\left(1+[(2k+1)n-2]\lambda-i\frac{n\lambda\theta}{\pi}\right)}\frac{\Gamma\left(1+2kn\lambda+i\frac{n\lambda\theta}{\pi}\right)}{\Gamma\left(1+2kn\lambda-i\frac{n\lambda\theta}{\pi}\right)}.
\end{multline}
This infinite product can be expressed very neatly with the help of the following integral representation
\begin{equation}
    \log\Gamma(z) = \int_0^\infty \frac{dt}{t}\left[(z-1)e^{-t}+\frac{e^{-tz}-e^{-t}}{1-e^{-t}}  \right].
\end{equation}
That results in our final answer
\begin{equation}
    \log F(\theta) = 2i\int_0^\infty \frac{\cosh\frac{\pi\omega (n-2)}{2n}\sinh\frac{\pi\omega}{n}(\frac{1}{2\lambda}-1)}{\cosh\frac{\pi\omega}{2}\sinh\frac{\pi\omega}{2n\lambda}}\sin\theta\omega\frac{d\omega}{\omega}. 
\end{equation}

\section{TBA}
In this appendix we are going to use the technique of Thermodynamic Bethe Ansatz to relate the parameters of the ATFT Lagrangian with the physical mass of fundamental particles present in our theory. 

\subsection{Bethe Ansatz equation}

The dual regime of the YB-deformed $\OO(2n)$ model has the unbroken $\left[\mathrm{U}(1)\right]^{n+1}$ global group, which allows for a total of $n+1$ conserved charges $Q_i=\int \psi^*_i\psi_i dx$. The usual way of finding ground state energy via TBA is to add the chemical potential term of the form $\mu\cdot Q$ to the Lagrangian, and then to take $\mu$ to infinity, which will result in the condensation of the corresponding particles. Since all the particles in the integrable multiplet are equivalent, it suffices to condense only the "first" fermion: $\mu_1=\mu$,  $\mu_{i\geq 2}=0$. In the large $\mu/m$ limit the massive term in the Lagrangian can be neglected, and the effective theory is the one-fermion massless Thirring model with action
\begin{equation}
S_{TM}=\int\left[\overline{\psi}_1\left(i\hat{\partial}+\hat{A}\right)\psi_1+\frac{1+b^2}{2\pi b^2}A^\nu A_\nu\right]d^2x = \int \left(i\overline{\psi}_1\hat{\partial}\psi_1+\frac{\pi b^2}{2(1+b^2)}\left(\overline{\psi}_1\gamma^\nu\psi_1\right)^2\right)d^2x
\end{equation}

We start off by calculating the ground state energy. In the ground state the particles fill all states inside the Fermi interval ($\theta\in(-B,B)$). The distribution function $\epsilon(\theta)$ is known to satisfy the Bethe Ansatz dressing equation
\begin{equation}
\mu - m\cosh\theta = \epsilon(\theta)-\int_{-B}^B K(\theta-\theta')\epsilon(\theta')d\theta'.
\end{equation}
Here the kernel $K(\theta)$ is related to the $11\rightarrow 11$ scattering phase $F(\theta)$ in the following way:
\begin{equation}
	K(\theta) = \frac{1}{2\pi}\frac{d}{d\theta}\delta(\theta) = \frac{1}{2\pi i}\frac{d}{d\theta}\left(\log F(\theta)\right)
\end{equation}
This distribution function should also obey $\epsilon(\pm B)=0$ and $\epsilon(\theta)=\epsilon(-\theta)$.

It is instructive to redefine the kernel in such a way that the \textit{r.h.s} becomes an integral operator acting on the distribution function 
\begin{equation}
\Tilde{K}(\theta)=\delta(\theta)-\frac{1}{2\pi i}\frac{d}{d\theta}\left(\log F(\theta)\right)
\end{equation}
In what follows we will also need the Fourier transform of this kernel:
\begin{equation}
K(\omega)=\int_{-\infty}^\infty \Tilde{K}(\theta)e^{i\omega \theta}d\theta = \frac{\sinh \frac{\pi \omega}{n}\cosh\left(\frac{\pi\omega}{n}\left[\frac{1}{2\lambda}+\frac{n}{2}-1\right]\right)}{\cosh\frac{\pi \omega}{2}\sinh\frac{\pi\omega}{2n\lambda}}
\end{equation}

In these notations the BA equations are
\begin{align}
\label{eq:DressedDistribution}
\varepsilon(\mu) = \varepsilon(0)-\frac{m}{2\pi}\int_{-B}^B\cosh \theta \,\epsilon (\theta)d\theta,\\
\label{eq:DressedGroundStateEnergy}
\mu  - m\cosh \theta = \int_{-B}^B\Tilde{K}(\theta-\theta')\epsilon(\theta')d\theta'.
\end{align}
$\varepsilon(\mu)$ is the ground state energy as a function of the chemical potential $\mu$.

\paragraph{UV limit} Consider the limit $\mu \to \infty$. Since in this limit the "Fermi rapidity" $B$ also tends to $\infty$, the main contribution to the integral $\int_{(-B,B)}\cosh\theta \epsilon(\theta)d\theta$ comes from the regions where the exponents are large enough, so we can set the other limit of integration to infinity:
\begin{equation}
\int_{-B}^B\cosh\theta\epsilon(\theta)d\theta \approx \frac{1}{2}\int_{-\infty}^Be^\theta\epsilon(\theta) d\theta+\frac{1}{2}\int_{-B}^\infty e^\theta\epsilon(\theta) d\theta = \int_{-\infty}^B e^{\theta}\frac{\epsilon(\theta)+\epsilon(-\theta)}{2}= \int_{-\infty}^Be^{\theta}\epsilon(\theta)d\theta,
\end{equation}
in the last equation we used that $\epsilon(\theta)=\epsilon(-\theta)$. Now we make the similar approximation $\cosh \theta \approx \frac{1}{2}e^\theta$ in the equation (\ref{eq:DressedDistribution}), which results in
\begin{equation}
\mu-\frac{1}{2}me^{\theta}=\int_{-\infty}^B\Tilde{K}(\theta-\theta')\epsilon(\theta')d\theta',
\end{equation}
which we can differentiate with respect to $\theta$ to then express the exponent $e^\theta$ and substitute it into the BA equation for the ground state energy
\begin{equation}
\varepsilon(\mu)=\varepsilon(0)+\frac{\mu}{2\pi}\int_{-\infty}^B\epsilon'(\theta)d\theta = \varepsilon(0)-\frac{\mu \epsilon(-\infty)}{2\pi}.
\end{equation}

\subsection{Wiener-Hopf method}
The equation (\ref{eq:DressedGroundStateEnergy}) can be solved by the modified WH method due to the specific property of the kernel $K(\omega)$: it can be represented as the product
\begin{equation}
	K(\omega)=\frac{1}{K_+(\omega)K_-(\omega)},
\end{equation}
where $K_+(\omega)$ is analytic in the upper half-plane and $K_-(\omega)=K_+(-\omega)$. Indeed,
\begin{equation}
K(\omega) = \frac{1}{2\lambda}\left[\frac{\Gamma(\frac{1}{2}+\frac{i\omega}{2})\Gamma(\frac{i\omega}{2n\lambda})}{\Gamma(\frac{i\omega}{n})\Gamma(\frac{1}{2}+\frac{i\omega}{n}[\frac{1}{2\lambda}+\frac{n}{2}-1])}\right]\times\left[\omega\mapsto -\omega\right].
\end{equation}
Thus 
\begin{equation}
K_+(\omega)=f(\omega)\sqrt{2\lambda}\frac{\Gamma\left(\frac{i\omega}{n})\Gamma(\frac{1}{2}+\frac{i\omega}{n}[\frac{1}{2\lambda}+\frac{n}{2}-1]\right)}{\Gamma\left(\frac{1}{2}+\frac{i\omega}{2}\right)\Gamma\left(\frac{i\omega}{2n\lambda}\right)},
\end{equation}
where $f(\omega)$ satisfies the property $f(\omega)f(-\omega)=1$ and can be extracted from the condition $K_+(\omega)=1+O(1/\omega)$ as $\omega \to \infty$ in any direction except positive imaginary semi-axis. That fixes \begin{equation}
	f(\omega)=e^{i\omega \Delta}, \quad \text{with}\quad \Delta = \frac{1}{n}\left[\frac{n}{2}\log\frac{n}{2}+\frac{1}{2\lambda}\log\frac{1}{2\lambda}+\left(1-\frac{n}{2}-\frac{1}{2\lambda}\right)\log\left(\frac{1}{2\lambda}+\frac{n}{2}-1\right)\right].
\end{equation}
It can be shown \cite{Zamolodchikov:1995xk} that the equation for the ground state energy reduces to
\begin{equation}
\label{eq:GroundStateEnergyWH}
\varepsilon(\mu)-\varepsilon(0)=-\frac{\mu^2 u(i)}{2\pi K(0)}\left[1-\int_{C_+'}\frac{e^{2i\omega' B}}{\omega'-i}\rho(\omega')u(\omega')\frac{d\omega'}{2\pi i} \right],
\end{equation}
where contour encircles all the poles of functions $\rho$ and $u$ in the upper half-plane and explicitly the pole in $\omega' = i$. Function $\rho(\omega)$ and $u(\omega)$ are defined as
\begin{gather}
u(\omega) = \frac{i}{\omega}+\int_{C_+}\frac{e^{2i\omega' B}}{\omega+\omega'}\rho(\omega')u(\omega')\frac{d\omega'}{2\pi i},\\
\rho(\omega) = \frac{1-i\omega}{1+i\omega}\frac{K_+(\omega)}{K_-(\omega)} = \frac{f(\omega)}{f(-\omega)}\frac{\Gamma(\frac{i\omega}{n})\Gamma(\frac{1}{2}+\frac{i\omega}{n}(\frac{1}{2\lambda}+\frac{n}{2}-1))}{\Gamma(\frac{3}{2}+\frac{i\omega}{2})\Gamma(\frac{i\omega}{2n\lambda})}\frac{\Gamma(\frac{3}{2}-\frac{i\omega}{2})\Gamma(\frac{-i\omega}{2n\lambda})}{\Gamma(\frac{-i\omega}{n})\Gamma(\frac{1}{2}-\frac{i\omega}{n}(\frac{1}{2\lambda}+\frac{n}{2}-1))} ,
\end{gather}
here contour $C_+$ encircles all the poles in the upper half-plane of the function $\rho$. In addition to the explicit pole at $\omega=i$, there are two series of them, namely, $\omega^{(1)}_k=ink$ and $\omega^{(2)}_k=i(2k-1)\frac{n\lambda}{\lambda(n-2)+1}$ with $k\in \mathbb{N}$. Also for convenience we will denote
\begin{equation}
	a_k=\frac{i}{n}e^{2kn\Delta}\underset{\omega^{(1)}_k}{\mathrm{res}\,}\rho(\omega)=\frac{(-1)^k}{k!(k-1)!}\frac{\Gamma(\frac{k}{2\lambda})\Gamma(\frac{3+nk}{2})\Gamma(\frac{1}{2}-k(\frac{1}{2\lambda}+\frac{n}{2}-1))}{\Gamma(-\frac{k}{2\lambda})\Gamma(\frac{3-nk}{2})\Gamma(\frac{1}{2}+k(\frac{1}{2\lambda}+\frac{n}{2}-1))}
\end{equation}
and
\begin{equation}
	b_k=e^{\frac{2n\lambda(2k-1)}{\lambda(n-2)+1}\Delta}\frac{\lambda(n-2)+1}{i2n\lambda}\underset{\omega^{(2)}_k}{\mathrm{res}\,}\rho(\omega)=\frac{(-1)^k}{((k-1)!)^2}\frac{\Gamma(-\frac{(2k-1)\lambda}{\lambda(n-2)+1})\Gamma(\frac{2\lambda(n(k+1)-3)+3}{2\lambda(n-2)+2})\Gamma(\frac{2k-1}{2\lambda(n-2)+2})}{\Gamma(\frac{(2k-1)\lambda}{\lambda(n-2)+1})\Gamma(\frac{3-2\lambda(n(k-2)+3}{2\lambda(n-2)+2})\Gamma(-\frac{2k-1}{2\lambda(n-2)+2})}
\end{equation}

The boundary conditions $\epsilon(\pm B)=0$ are equivalent to
\begin{gather}
i\mu K_+(0)-\frac{iMe^B}{2}K_+(-i)=-i\mu K_+(0)\int_{C_+}\frac{e^{2i\omega B}}{\omega +i}\rho(\omega)u(\omega)\frac{d\omega}{2\pi i} = -i\mu K_+(0)(u(i)-1)\nonumber \\ u(i) = \frac{M e^B}{2\mu}\frac{K_+(-i)}{K_+(0)} = \frac{\lambda Me^{B+\Delta}}{\mu}\frac{\Gamma(\frac{1}{n})\Gamma(1+\frac{1}{2n\lambda}-\frac{1}{n})}{\Gamma(\frac{1}{2n\lambda})} = y e^{B+\Delta},
\end{gather}
with
\begin{equation}
y=\frac{\lambda M}{\mu}\frac{\Gamma(\frac{1}{n})\Gamma(1+\frac{1}{2n\lambda}-\frac{1}{n})}{\Gamma(\frac{1}{2n\lambda})}.
\label{eq:yParameter}
\end{equation}

We can use the expression (\ref{eq:GroundStateEnergyWH}) to obtain the series expansion with $B\to \infty$ as perturbation parameter.
\begin{equation}
u(i) = 1-\sum_{k=1}^\infty\frac{1}{kn+1}q^nw_ka_k+\sum_{k=1}^\infty\frac{\lambda(n-2)+1}{(2k-1)n\lambda+\lambda(n-2)+1}p^{2k-1}u_kb_k
\end{equation}
where 
\begin{gather}
q=e^{-2n(B+\Delta)}\quad \text{and} \quad w_k=n\cdot u(ikn),\\
p=\exp\left(-\frac{2\lambda n}{\lambda(n-2)+1}(B+\Delta)\right)\quad \text{and} \quad u_k = \frac{2n\lambda}{\lambda(n-2)+1}u\left(\frac{i(2k-1)\lambda n}{\lambda(n-2)+1}\right)
\end{gather}

The similar calculation in the integral in the equation (\ref{eq:GroundStateEnergyWH}) result into the final expression 
\begin{multline}
\varepsilon(\mu)-\varepsilon(0)=-\frac{\mu^2}{2\pi K(0)}\left(1-\sum_{k=1}^\infty\frac{1}{kn+1}q^nw_ka_k+\sum_{k=1}^\infty\frac{\lambda(n-2)+1}{(2k-1)n\lambda+\lambda(n-2)+1}p^{2k-1}u_kb_k\right)\times\nonumber \\ \times \left(1-\lim_{\omega'\to i}\frac{d}{d\omega'}\left[(\omega'-i)^2e^{2i\omega' B}\rho(\omega')u(\omega')\right]+\sum_{k=1}^\infty\frac{1}{kn-1}q^nw_ka_k-\sum_{k=1}^\infty\frac{\lambda(n-2)+1}{(2k-1)n\lambda-\lambda(n-2)-1}p^{2k-1}u_kb_k \right)
\end{multline}
The limit can be simplified to
\begin{multline}
\lim_{\omega'\to i}\frac{d}{d\omega'}\left[(\omega'-i)e^{2i\omega' B}\rho(\omega')u(\omega') \right]=-u(i)e^{-2B}\lim_{\omega'\to i}\frac{d}{d\omega'}\left[(\omega'+i)\frac{K_+(\omega')}{K_-(\omega')}\right]=\\= -\frac{i}{2}u(i)e^{-2B}\lim_{\omega'\to i}\left[(\omega'+i)\frac{K_+(\omega')}{K_-(\omega')}\psi\left(\frac{1+i\omega'}{2}\right)\right]
\end{multline}

In the first two orders in $\sim e^{-B}$ we get 
\begin{gather}
\varepsilon(\mu)-\varepsilon(0) = -\frac{\mu^2}{2\pi K(0)}\left(1-\frac{i}{2}\lim_{\omega'\to i}\left[(\omega'+i)\frac{K_+(\omega')}{K_-(\omega')}\psi\left(\frac{1+i\omega'}{2}\right)\right]e^{-2B}+O(e^{-2B})\right),
\end{gather}
where $\varepsilon(0)$ corresponds exactly to the pole in $\omega = i$ \cite{Zamolodchikov:1995xk}.
\begin{multline}
\underset{\omega=i}{\mathrm{res}}\frac{1}{\omega-i}\rho(\omega) = \lim_{\omega\to i}\frac{d}{d\omega}\left[(\omega-i)^2\frac{1}{\omega-i}\frac{1-i\omega}{1+i\omega}\frac{K_+(\omega)}{K_-(\omega)}\right]=\\
=-\frac{1}{K_-(i)}\lim_{\omega\to i}\frac{d}{d\omega}\left[(\omega+i)K_+(\omega)\right] = \frac{\Gamma(-\frac{1}{n})\Gamma(\frac{1}{n}-\frac{1}{2n\lambda})\Gamma(\frac{1}{2n\lambda})}{\Gamma(\frac{1}{n})\Gamma(1-\frac{1}{n}+\frac{1}{2n\lambda})\Gamma(-\frac{1}{2n\lambda})}e^{-2\Delta}.
\end{multline}
For $\varepsilon(0)$ we have
\begin{multline}
\varepsilon(0)=\frac{\mu^2u(i)^2}{2\pi K(0)}e^{-2B}\underset{\omega=i}{\mathrm{res}}\frac{1}{\omega-i}\rho(\omega) = \nonumber\\\nonumber =\frac{\mu^2}{2\pi K(0)}e^{-2B}\frac{\lambda^2 M^2 e^{2B+2\Delta}}{\mu^2}\frac{\Gamma(\frac{1}{n})\Gamma(1+\frac{1}{2n\lambda}-\frac{1}{n})}{\Gamma(\frac{1}{2n\lambda})} e^{-2\Delta}\frac{\Gamma(-\frac{1}{n})\Gamma(\frac{1}{n}-\frac{1}{2n\lambda})}{\Gamma(-\frac{1}{2n\lambda})} = \\ =\frac{\mu^2}{2\pi K(0)}e^{-2B}\frac{\lambda^2 M^2 e^{2B}}{\mu^2}\frac{\pi}{2\lambda}\frac{\sin\frac{\pi}{2n\lambda}}{\sin\frac{\pi}{n}\sin\frac{\pi (2\lambda -1)}{2n\lambda}} =\frac{M^2\sin\frac{\pi}{2n\lambda}}{8\sin\frac{\pi}{n}\sin\frac{\pi (2\lambda -1)}{2n\lambda}},
\end{multline}
And remaining part corresponds to $\varepsilon(\mu)$:
\begin{multline}
\varepsilon(\mu)=-\frac{\mu^2}{2\pi K(0)}\left(1-\sum_{k=1}^\infty\frac{1}{kn+1}q^kw_ka_k+\sum_{k=1}^\infty\frac{\lambda(n-2)+1}{(2k-1)n\lambda+\lambda(n-2)+1}p^{2k-1}u_kb_k\right)\times\nonumber \\ \times \left(1+\sum_{k=1}^\infty\frac{1}{kn-1}q^kw_ka_k-\sum_{k=1}^\infty\frac{\lambda(n-2)+1}{(2k-1)n\lambda-\lambda(n-2)-1}p^{2k-1}u_kb_k \right).
\end{multline}

We wish to expand the ground state energy in powers of $M/\mu$ to compare it with the perturbative results. To do so we use the $y$ parameter which entered in the boundary condition (\ref{eq:yParameter}). It can be rewritten as
\begin{equation}
	y = e^{-(B+\Delta)}\left(1-\sum_{k=1}^\infty\frac{1}{kn+1}q^kw_ka_k+\sum_{k=1}^\infty\frac{\lambda(n-2)+1}{(2k-1)n\lambda+\lambda(n-2)+1}p^{2k-1}u_kb_k\right).
\end{equation}
We expand $\varepsilon(\mu)$ in powers of $y$:
\begin{equation}
\varepsilon(\mu) = -\frac{\mu^2}{2\pi K(0)}\sum_{k=0}^\infty \alpha_k y^k.
\end{equation}
The first coefficient in this expansion reads
\begin{gather}
\alpha_1=\frac{-2}{n^2-1}\frac{\Gamma\left(\frac{1}{2\lambda}\right)\Gamma\left(\frac{3}{2}-\frac{1}{2\lambda}-\frac{n}{2}\right)\Gamma\left(\frac{3}{2}+\frac{n}{2}\right)}{\Gamma\left(-\frac{1}{2\lambda}\right)\Gamma\left(\frac{1}{2\lambda}+\frac{n}{2}-\frac{1}{2}\right)\Gamma\left(\frac{3}{2}-\frac{n}{2}\right)}y^{2n}.
\end{gather}

\subsection{Perturbative calculation}
Next, we compute $\alpha_1$ perturbatively and compare the results. Standard QFT procedure is to treat the partition function as a formal functional integral $Z = \int e^{-S_\text{ATFT}[\phi]}[D\phi]$, where $S_\mathrm{ATFT} = S_\mathrm{free} + \Lambda S_\mathrm{int}$. One then expands $e^{-S_\mathrm{ATFT}}$ in powers of $\Lambda$ and computes correlation order by order in $\Lambda$. We start with the ATFT Lagrangian of the dual regime \eqref{ATFT-dual-general}. For the $\OO(2n)$ dual theory the first nontrivial contribution to $Z = \langle 1 \rangle$ reads
\begin{align}
\langle 1\rangle = \langle 1\rangle_0+\frac{1}{(4(n-1))!}(-\Lambda)^{4(n-1)}\int\left\langle e^{(\boldsymbol{\alpha}_1,\boldsymbol{\phi}(x_1))}e^{(\boldsymbol{\alpha}_2,\boldsymbol{\phi}(x_2))}e^{(\boldsymbol{\alpha}_3,\boldsymbol{\phi}(x_3))}e^{(\boldsymbol{\alpha}_3,\boldsymbol{\phi}(x_4))}\cdot\dots \right.\times\\\times\left. e^{(\boldsymbol{\alpha}_{2(n-1)},\boldsymbol{\phi}(x_{4(n-1)-3}))} e^{(\boldsymbol{\alpha}_{2(n-1)},\boldsymbol{\phi}(x_{4(n-1)-2}))}e^{(\boldsymbol{\alpha}_{2N-1},\boldsymbol{\phi}(x_{4(n-1)-1}))}e^{(\boldsymbol{\alpha}_{2N},\boldsymbol{\phi}(x_{4(n-1)})}  \right\rangle_0\times \\ \times e^{i\beta\mu(x_1^1+x_2^1-x_{4(n-1)-1}^1-x_{4(n-1)}^1)}\prod_{i=1}^{n-2}e^{2i\beta \mu (x_{2i+1^1}-x_{2i}^1)} \prod_{i=1}^{4(n-1)}d^2x_i
\end{align}
For $n = 2$, that is, for dual $\OO(4)$ model we have
\begin{multline}
    \langle1\rangle-\langle1\rangle_0 =\\
    = \frac{(-\Lambda)^4}{4!}4!\int \left\langle e^{-b\Phi_1(x_1)+i\beta \phi(x_1)}e^{b\Phi_1(x_2)+i\beta \phi(x_2)}e^{b\Phi_2(x_3)-i\beta \phi(x_3)}e^{-b\Phi_2(x_4)-i\beta \phi(x_4)} \right\rangle_0 e^{i\beta\mu(x_1^1+x_2^1-x_3^1-x_4^1)}\prod_{i=1}^4d^2x_i =\\
    =\frac{(-\Lambda)^4}{(\beta \mu)^{6+4(b^2-\beta^2)}}\int d^2x_4\int \frac{|x_1-x_2|^{2(b^2+\beta^2)}|x_3|^{2(b^2+\beta^2)}}{|x_1-x_3|^{2\beta^2}|x_1|^{2\beta^2}|x_2-x_3|^{2\beta^2}|x_2|^{2\beta^2}}e^{i(x_1^1+x_2^1-x_3^1)}\prod_{i=1}^3d^2x_i
\end{multline}
in the last equality we made a linear transformation $x_i \to \frac{1}{\beta\mu}\left(x_i+x_4 \right)$. To evaluate the innermost integral it is useful to treat the integration against $e^{i(x_1^1 + x_2^1 - x_3^1)}$ as the Fourier transformation. So, we perform yet another linear change of coordinates
\begin{equation}
    \left(
    \begin{matrix}
        x_1^j\\x_2^j\\x_3^j\\x_4^j
    \end{matrix}
    \right) = \left(
    \begin{matrix}
        1&1&-1&0\\
        1&-1&1&0\\
        -1&1&1&0\\
        0&0&0&1
    \end{matrix} \right)
    \left(
    \begin{matrix}
        u^j\\v^j\\p^j\\q^j
    \end{matrix}\right),
\end{equation}
where index $j=0,1$ labels the temporal and spatial coordinates. The determinant of this $8\times 8$ matrix is $2^4$. In these new coordinates the integral in question is
\begin{gather}
    I_4 = 2^4\int d^2u\,d^2v\,d^2p\  e^{iu^1}|v-p|^{2(b^2+\beta^2)}|v+p|^{2(b^2+\beta^2)}|u+v|^{-2\beta^2}|u-v|^{-2\beta^2}|u+p|^{-2\beta^2}|u-p|^{-2\beta^2} = \\ = 2^4\frac{1}{4}\int d^2u\,e^{iu^1}\int\,d^2v\,d^2p |v^2-p^2|^{2(b^2-\beta^2)}|v^2-u^2|^{-2\beta^2}|p^2-u^2|^{-2\beta^2}.
\end{gather}
In the last equation we replaced $\{x,y\}\to \{z,\overline{z}\}$ in $t$ and $s$ variables, which results in a factor of $-1/4$. Next, we change the variables $t=v^2$ and $s=p^2$ and rearrange the integrations:
\begin{equation}
I_4 = -2^4\frac{4}{2^4}\frac{1}{4}\int d^2u\, e^{iu^1}|u|^{4(1+b^2-\beta^2)}\int d^2t\,d^2s\, |t|^{-1}|s|^{-1}|t-1|^{-2\beta^2}|s-1|^{-2\beta^2}|t-s|^{2(b^2+\beta^2)}.
\end{equation}
The inner integral is the standard Selberg integral. Evaluating it and the outer $u$-integral without specifying the dependence between $b^2$ and $\beta^2$, we obtain the following result:
\begin{gather*}
    \int d^2u\, e^{iu^1} |u|^{4(1+b^2-\beta^2)} = -2\sqrt{\pi}\cos[2\pi (b^2-\beta^2)] \frac{\Gamma(6+4[b^2-\beta^2])\Gamma(-\frac{5}{2}-2b^2+2\beta^2)}{\Gamma(2[\beta^2-b^2-1])} \\
    \int d^2t\,d^2s\, |t|^{-1}|s|^{-1}|t-1|^{-2\beta^2}|s-1|^{-2\beta^2}|t-s|^{2(b^2+\beta^2)} = 2I_2\left(-\frac{1}{2},-\beta^2|\frac{b^2+\beta^2}{2}\right) = \\ 2\pi^2\frac{\gamma\left(\frac{1}{2}\right)  \gamma\left(1-\beta^2\right)}{\gamma\left(\frac{3}{2}-\beta^2+\frac{b^2+\beta^2}{2}\right)}\frac{ \gamma\left(b^2+\beta^2\right)}{\gamma\left(\frac{b^2+\beta^2}{2}\right)}\frac{\gamma\left(\frac{1}{2}+\frac{b^2+\beta^2}{2}\right)  \gamma\left(1-\beta^2+\frac{b^2+\beta^2}{2}\right)}{\gamma\left(\frac{3}{2}-\beta^2+b^2+\beta^2\right)},
\end{gather*}
where $\gamma(x) = \frac{\Gamma(x)}{\Gamma(1-x)}$. Multiplying this functions of $\beta^2$ and $b^2$ and demanding the relation $\beta^2=1+b^2$ we obtain the result for the first perturbative term to the $Z$
\begin{equation}
    \langle 1\rangle-\langle 1\rangle_0 = -\frac{(-\Lambda)^4}{(1+b^2)^2\mu^2}\frac{2^{2+4(1+b^2)}\pi^3\Gamma(-\frac{1}{2}-b^2)\Gamma(1+b^2)}{\Gamma(-b^2)\Gamma(\frac{3}{2}+b^2)} = -\frac{\Lambda^4}{\mu^2}(2\lambda)^2\frac{2^{2+\frac{2}{\lambda}}\pi^3\Gamma(\frac{1}{2\lambda})\Gamma(\frac{1}{2}-\frac{1}{2\lambda})}{\Gamma(1-\frac{1}{2\lambda})\Gamma(\frac{1}{2}+\frac{1}{2\lambda})}
\end{equation}


Comparing this term to $-\frac{\mu^2}{4\pi\lambda}\alpha_1$, we extract the relation between the coupling constant $m = 4\pi \Lambda$ and the physical mass $M$:
\begin{gather}
    M = m \frac{2^{\frac{1}{2\lambda}}\Gamma\left(\frac{1}{4\lambda}\right)}{\Gamma\left(\frac{1}{2}\right)\Gamma\left(\frac{1}{2}+\frac{1}{4\lambda}\right)}
\end{gather}
This result is quite similar to the one obtained in \cite{Fateev:2018yos} for the non-twisted version $D_n^{(1)}$. The difference is in the power of $2$ which arguably can be explained in the perturbation theory approach, where we replace $b^2$ with $-\beta^2 = -\frac{1}{2\lambda}$ in the twisted regime. The structure of $\Gamma$-functions coming from BA calculations coincide exactly.

\printbibliography

\end{document}